%% file: paper.tex
\documentclass{aa}
\usepackage{graphicx}
\usepackage{bm}
\usepackage{times}
\usepackage{amsmath}
\usepackage{amssymb}
\usepackage{natbib}
\usepackage{color}
\usepackage{mathrsfs}
\usepackage[colorlinks,allcolors=blue]{hyperref}
\usepackage{url}
\usepackage{multirow}
\bibpunct{(}{)}{;}{a}{}{,}
\graphicspath{{./fig/}{./png/}}
\input macros

\begin{document}

\authorrunning{K\"apyl\"a}
\titlerunning{Convective scale and subadiabatic layers in rotating convection}

   \title{Convective scale and subadiabatic layers in\\ simulations of rotating compressible convection}

   \author{P. J. K\"apyl\"a
          \inst{1,2}
          }

   \institute{Leibniz-Institute for Solar Physics (KIS), Sch\"oneckstr.\ 6, D-79104 Freiburg, Germany 
              email: \href{mailto:pkaepyl@uni-goettingen.de}{pkapyla@leibniz-kis.de}
   \and
              Georg-August-University G\"ottingen, Institute for Astrophysics and Geophysics,
              Friedrich-Hund-Platz 1, D-37077 G\"ottingen, Germany
}

\date{\today}

\abstract{%
    Rotation is thought to influence the size of convective eddies and
    the efficiency of convective energy transport in the deep
    convection zones of stars. Rotationally constrained convection has
    been invoked to explain the lack of large-scale power in
    observations of solar flows.
   }%
   {The main aims are to quantify the effects of rotation on the scale
     of convective eddies and velocity, the depths of convective
     overshoot, and the subadiabatic Deardorff layers.
   }%
   {Moderately turbulent three-dimensional hydrodynamic simulations of
     rotating convection in local Cartesian domains were run. The
     rotation rate and luminosity of the simulations are varied to
     probe the dependency of the results on Coriolis, Mach, and
     Richardsson numbers measuring the incluences of rotation,
     compressibility, and stiffness of the radiative layer. The
     results were compared with theoretical scaling results that
     assume a balance between Coriolis, inertial, and buoyancy
     (Archemedean) forces, which is also referred to as the CIA
     balance.
   }%
   {The horizontal scale of convective eddies decreases as rotation
     increases, and ultimately reaches a rotationally constrained
     regime consistent with the CIA balance. Using a new measure of
     the rotational influence on the system, it is shown that even the
     deep parts of the solar convection zone are not in the
     rotationally constrained regime. The simulations capture the
     slowly and rapidly rotating scaling laws predicted by theory, and
     the Sun appears to be in between these two regimes. Both, the
     overshooting depth and the extent of the Deardorff layer,
     decrease as rotation becomes more rapid. For sufficiently rapid
     rotation the Deardorff layer is absent due to the symmetrization
     of up- and downflows. However, for the most rapidly rotating
     cases the overshooting increases again due to unrealistically
     large Richardsson numbers that allow convective columns penetrate
     deep into the radiative layer.
   }%
   {Relating the simulations with the Sun suggests that the convective
     scale even in the deep parts of the Sun is only mildly affected
     by rotation and that some other mechanism is needed to explain
     the lack of strong large-scale flows in the Sun. Taking the
     current results at face value, the overshoot and Deardorff layers
     are estimated to span roughly five per cent of the pressure scale
     height at the base of the convection zone in the Sun.
   }%

   \keywords{   turbulence -- convection
   }

  \maketitle


\section{Introduction}

The theoretical understanding of solar and stellar convection was
shaken roughly a decade ago when helioseismic analysis suggested that
the velocity amplitudes in the deep solar convection zone are orders
of magnitude smaller than anticipated from theoretical and numerical
models \citep{HDS12}. Significant effort has been put in refining
these estimates but a gaping discrepancy between numerical models and
helioseismology remains
\cite[e.g.][]{2016AnRFM..48..191H,2021PhDT........26P}; see, however
\cite{GHFT15}. This issue is now refererred to as the convective
conundrum \citep{2016AdSpR..58.1475O}.

Several solutions to this conundrum have been proposed, including high
effective Prandtl number \citep[e.g.][]{2018PhFl...30d6602K},
rotationally constrained convection \citep{FH16b}, and that the
superadiabatic layer in the Sun is much thinner than thus far thought
\citep{Br16}; see also \cite{2023SSRv..219...58K} and references
therein. The two latter ideas are explored further in the current
study. The idea that convection is rotationally constrained in the
deep convection zone (CZ) is already borne out of mixing length models
of solar convection that imply velocity amplitudes $\uconv$ of about
$10$~m~s$^{-1}$ for the deep convection zone, while the convective
length scale $\lconv$, which is the mixing length, is of the order of
$100$~Mm, yielding a Coriolis number $\Co_\odot=2\Omsun\lconv/\uconv$
of the order of 10 \citep[e.g.][]{O03,2020RvMP...92d1001S}. However,
this estimate does not take into account the decreasing length scale
due to the rotational influence on convection. Assuming that the
Coriolis, inertial, and buoyancy (Archimedean) forces balance, also
known as the CIA balance
\citep[e.g.][]{1979GApFD..12..139S,1982Icar...52...62I,2013E&PSL.371..156K,2014ApJ...791...13B,2020PhRvR...2d3115A,2021PNAS..11822518V},
implies that the convective scale is given by $\lconv \propto
\Co^{-1/2}$, where $\Co=2\Omega H/\uconv$ is a global Coriolis number,
where $\Omega$ is the rotation rate and where $H$ is a length scale
corresponding to the system size
\citep[e.g.][]{2020PhRvR...2d3115A}. This idea has been explored
recently by \cite{FH16b} and \cite{2021PNAS..11822518V} who suggested
that the largest convectively driven scale in the Sun coincides with
that of supergranulation due to rotationally constrained convection in
the deep CZ. These studies assumed from the outset that convection is
strongly rotationally affected. Here a somewhat different perspective
is taken in that an attempt is made to assess whether this assumption
holds for the deep solar CZ. Furthermore, in addition to $\lconv$, the
scalings of various quantities based on predictions from the CIA
balance are studied over a wide range of rotation rates.

Simulations of stratified overshooting convection have revealed that
deep parts of CZs are often weakly stably stratified
\citep[e.g.][]{1993A&A...277...93R,2015ApJ...799..142T,2017ApJ...843...52H,2017ApJ...851...74B,2017ApJ...845L..23K,2019A&A...631A.122K}. This
is interpreted such that convection is driven by the cooling at the
surface that induces cool downflow plumes which pierce through the
entire convection zone and penetrate deep into the stable layers
below. This process has been named entropy rain \citep[e.g.][]{Br16}
and goes back to ideas presented by \cite{Sp97} and the simulations of
Stein and Nordlund \citep[e.g.][]{SN89,1998ApJ...499..914S}. This
picture of convection is a clean break from the canonical view in
which convection is driven throughout the convection zone by a
superadiabatic temperature gradient, an idea which is also encoded
into the mixing length concept
\citep[e.g.][]{Vi53,BV58}. Theoretically this can be understood such
that the convective energy flux that is traditionally proportional to
the entropy gradient is supplemented by a non-gradient term
proportional to the variance of entropy fluctuations
\citep{1961JAtS...18..540D,De66}.

Analysis of the force balance of up- and downflows in non-rotating
hydrodynamic simulations supports the idea of surface-driven non-local
convection
\citep[e.g.][]{2017ApJ...845L..23K,2019A&A...631A.122K,2021A&A...655A..78K}.
Thus far these studies have mostly concentrated on non-rotating
convection \citep[see,
  however][]{2019GApFD.113..149K,2021A&A...645A.141V}. Here rotation
is included to study its impact on the formation and extent of stably
stratified Deardorff layers where the convective flux runs counter to
the entropy gradient. Another aspect of interest in astrophysics is
convective overshooting \citep[see, e.g.][for a recent
  review]{2023Galax..11...56A}. Numerical studies targeting
specifically overshooting have largely concentrated on non-rotating
cases
\citep[e.g.][]{1995A&A...295..703S,1998A&A...340..178S,2000ApJ...529..402S,BCT02,2017ApJ...843...52H,2019A&A...631A.122K,2022ApJ...926..169A},
and the effects of rotation have received much less attention
\citep[e.g.][]{2003A&A...401..433Z,KKT04,2017ApJ...836..192B}. It is
generally thought that rotation leads to reduction of overshooting
depth \citep[e.g.][]{2003A&A...401..433Z} but a comprehensive study of
this is still lacking.

The remainder of the paper is organized as follows: the model is
described in \Sec{sect:model}, whereas the results and conclusions of
the study are presented in \Secs{sect:results}{sect:conclusions},
respectively. The derivations related to the CIA balance are presented
in \Appendix{app:convsca}.

\section{The model} \label{sect:model}

The model is the same as that used in
\cite{2019A&A...631A.122K,2021A&A...655A..78K}. The {\sc Pencil Code}
\citep[][]{2021JOSS....6.2807P}\footnote{\href{https://github.com/pencil-code/}{https://github.com/pencil-code/}}
was used to produce the simulations. Convection is modeled in a
Cartesian box with dimensions $(L_x,L_y,L_z)=(4,4,1.5)d$, where $d$ is
the depth of the initially convectively unstable layer. The equations
for compressible hydrodynamics are solved:
\begin{eqnarray}
\frac{D \ln \rho}{D t} &=& -\bm\nabla \bm\cdot \uuu, \label{equ:dens}\\
\frac{D\uuu}{D t} &=& {\bm g} -\frac{1}{\rho}(\bm\nabla p - \bm\nabla \bm\cdot 2 \nu \rho \bm{\mathsf{S}}) - 2\bm\Omega \times \uuu,\label{equ:mom} \\
T \frac{D s}{D t} &=& -\frac{1}{\rho} \left[\bm\nabla \bm\cdot \left(\FFF_{\rm rad} + \FFF_{\rm SGS}\right) - \mathcal{C} \right] + 2 \nu \bm{\mathsf{S}}^2,
\label{equ:ent}
\end{eqnarray}
where $D/Dt = \pd/\pd t + \uuu\cdot\bm\nabla$ is the advective
derivative, $\rho$ is the density, $\uuu$ is the velocity,
$\bm{g}=-g\hat{\bm{e}}_z$ is the acceleration due to gravity with
$g>0$, $p$ is the pressure, $T$ is the temperature, $s$ is the
specific entropy, $\nu$ is the constant kinematic viscosity, and
$\bm\Omega = \Omega_0(-\sin\theta,0,\cos\theta)^T$ is the rotation
vector, where $\theta$ is the colatitude. $\FFF_{\rm rad}$ and
$\FFF_{\rm SGS}$ are the radiative and turbulent subgrid scale (SGS)
fluxes, respectively, and $\mathcal{C}$ describes cooling near the
surface. $\SSt$ is the traceless rate-of-strain tensor with
\begin{eqnarray}
\SStij = \onehalf (u_{i,j} + u_{j,i}) - \onethird \delta_{ij} \bm\nabla\bm\cdot\uuu.
\end{eqnarray}
The gas is assumed to be optically thick and fully ionized, where
radiation is modeled via the diffusion approximation. The ideal gas
equation of state $p= (\cP - \cV) \rho T =\calR \rho T$ applies, where
$\calR$ is the gas constant, and $c_{\rm P}$ and $c_{\rm V}$ are the
specific heats at constant pressure and volume, respectively. The
radiative flux is given by
\begin{eqnarray}
\FFF_{\rm rad} = -K\bm\nabla T,
\label{equ:Frad}
\end{eqnarray}
where $K$ is the radiative heat conductivity
\begin{eqnarray}
K = \frac{16 \sigma_{\rm SB} T^3}{3 \kappa \rho},
\label{equ:Krad1}
\end{eqnarray}
where $\sigma_{\rm SB}$ is the Stefan-Boltzmann constant and $\kappa$
is the opacity. Assuming that the opacity is a power law of the form
$\kappa =\kappa_0 (\rho/\rho_0)^a (T/T_0)^b$, where $\rho_0$ and $T_0$
are reference values of density and temperature, the heat conductivity
is
\begin{eqnarray}
K(\rho,T) = K_0 (\rho/\rho_0)^{-(a+1)} (T/T_0)^{3-b}.
\label{equ:Krad2}
\end{eqnarray}
The choice $a=1$ and $b=-7/2$ corresponds to the Kramers opacity law
\citep{WHTR04}, which was used in convection simulations by
\cite{1990MNRAS.242..224E} and \cite{2000gac..conf...85B}.

Additional turbulent SGS diffusivity is applied for the entropy
fluctuations with
\begin{eqnarray}
\FFF_{\rm SGS} = -\rho T \chiSGS \bm\nabla s',
\label{equ:FSGS}
\end{eqnarray}
where $s'(\xxx)=s(\xxx)-\mean{s}$ with the overbar indicating
horizontal averaging. The coefficient $\chiSGS$ is constant in the
whole domain and $\FFF_{\rm SGS}$ has a negligible contribution to the
net energy flux such that $\mean{\FFF}_{\rm SGS} \approx 0$.

The cooling at the surface is described by
\begin{eqnarray}
\mathcal{C} = \rho\cP \frac{T_{\rm cool} - T}{\taucool} f_{\rm cool}(z),
\label{equ:cool}
\end{eqnarray}
where $\taucool$ is a cooling time, $T=e/c_{\rm V}$ is the temperature
where $e$ is the internal energy, and where $T_{\rm cool}=T_{\rm top}$
is a reference temperature corresponding to the fixed value at the top
boundary.

The advective terms in \Equs{equ:dens}{equ:ent} are written in terms
of a fifth-order upwinding derivative with a hyperdiffusive
sixth-order correction with a local flow-dependent diffusion
coefficient; see Appendix~B of \cite{DSB06}.

\subsection{Geometry, initial and boundary conditions}

The computational domain is a rectangular box where the vertical
coordinate is $z_{\rm bot} \leq z \leq z_{\rm top}$ with $z_{\rm
  bot}/d=-0.45$, $z_{\rm top}/d=1.05$. The horizontal coordinates $x$
and $y$ run from $-2d$ to $2d$. The initial stratification consists of
three layers. The two lower layers are polytropic with polytropic
indices $n_1=3.25$ ($z_{\rm bot}/d \leq z/d \leq 0$) and $n_2=1.5$ ($0
\leq z/d \leq 1$). The former follows from a radiative solution that
is a polytrope with index $n=(3-b)/(1+a)$; see \cite{BB14}, Appendix~A
of \cite{Br16}, and \Fig{fig:plot_hs}. The latter corresponds to a
marginally stable isentropic stratification. Initially the uppermost
layer above $z/d=1$ is isothermal, mimicking a photosphere where
radiative cooling is efficient. Convection ensues because the system
is not in thermal equilibrium due to the cooling near the surface and
due to the inefficient radiative diffusion in the layers above
$z/d=0$.  The velocity field is initially seeded with small-scale
Gaussian noise with amplitude $10^{-5}\sqrt{dg}$.

The horizontal boundaries are periodic and the vertical boundaries
are impenetrable and stress free according to
\begin{eqnarray}
\frac{\pd u_x}{\pd z} = \frac{\pd u_y}{\pd z} = u_z = 0.
\end{eqnarray}
A constant energy flux is imposed at the lower boundary by setting
\begin{eqnarray}
\frac{\pd T}{\pd z} = -\frac{F_\tbot}{K_\tbot},
\end{eqnarray}
where $F_{\rm bot}$ is the fixed input flux and
$K_\tbot=K(x,y,\zbot)$. Constant temperature $T=T_{\rm top}$ is
imposed on the upper vertical boundary.

\subsection{Units and control parameters}

The units of length, time, density, and entropy are given by
\begin{eqnarray}
[x] = d,\ \ \ [t] = \sqrt{d/g},\ \ \ [\rho] = \rho_0,\ \ \ [s] = \cP,
\end{eqnarray}
where $\rho_0$ is the initial value of density at $z=z_{\rm top}$. The
models are fully defined by choosing the values of $\nu$, $\Omega_0$,
$\theta$ $g$, $a$, $b$, $K_0$, $\rho_0$, $T_0$, $\taucool$, and the
SGS Prandtl number
\begin{eqnarray}
\PraSGS = \frac{\nu}{\chiSGS},
\end{eqnarray}
along with the cooling profile $f_{\rm cool}(z)$. The values of $K_0$,
$\rho_0$, $T_0$ are subsumed into another constant
$\widetilde{K}_0=K_0 \rho_0^{a+1} T_0^{b-3}$ which is fixed by
assuming the radiative flux at $\zbot$ to equal $\Fbot$ at $t=0$. The
cooling profile $f_{\rm cool}(z)=1$ above $z/d=1$ and $f_{\rm
  cool}(z)=0$ below $z/d=1$, connecting smoothly across the interface
over a width of $0.025d$. The quantity $\xi_0=\Hp^{\rm top}/d =
\mathcal{R}T_{\rm top}/gd$ sets the initial pressure scale height at
the surface and determining the initial density stratification. All of
the current simulations have $\xi_0=0.054$.

Prandtl number based on the radiative heat conductivity is
\begin{eqnarray}
  \Pr(\xxx) = \frac{\nu}{\chi(\xxx)},\label{equ:Pra}
\end{eqnarray}
where $\chi(\xxx)=K(\xxx)/\cP \rho(\xxx)$, quantifies the relative
importance of viscous to temperature diffusion. Unlike many other
simulations, $\Pr$ is not an input parameter because of the non-linear
dependence of the radiative diffusivity on the ambient thermodynamics.
The dimensionless normalized flux is given by
\begin{eqnarray}
  \Fn = \frac{\Fbot}{\rho(\zbot) c_{\rm s}^3(\zbot)},
\end{eqnarray}
where $\rho(\zbot)$ and $c_{\rm s}(\zbot)$ are the density and the
sound speed, respectively, at $z=z_{\rm bot}$ at $t=0$. At the base of
the solar CZ $\Fn\approx4\cdot10^{-11}$ \citep[e.g.][]{BCNS05},
whereas in the current fully compressible simulations several orders
of magnitude larger values are used.

The effect of rotation is quantified by the Taylor number
\begin{eqnarray}
\Ta = \frac{4\Omega_0^2 d^4}{\nu^2},
\end{eqnarray}
which is related to the Ekman number via $\Ek = \Ta^{-1/2}$.

The Rayleigh number based on the energy flux is given by
\begin{eqnarray}
\RaF = \frac{gd^4 \Fbot}{\cP \rho T \nu\chi^2}.
\end{eqnarray}
This can be used to construct a flux-based diffusion-free modified
Rayleigh number \citep[e.g.][]{2002JFM...470..115C,CA06}
\begin{eqnarray}
\RaFS = \frac{\RaF}{\Pra^2 \Ta^{3/2}},\label{equ:RaFS}
\end{eqnarray}
In the current set-up $\RaFS$ is given by
\begin{eqnarray}
\RaFS = \frac{g\Fbot}{8\cP\rho T \Omega_0^3 d^2}.
\end{eqnarray}
A reference depth needs to be chosen because
$\RaFS=\RaFS(z)$. Furthermore, $H \equiv \cP T/g$ is a length scale
related to the pressure scale height. The choice $d=H=\Hp$, where
$\Hp\equiv-(\pd\ln p/\pd z)^{-1}$ is the pressure scale height at the
base of the convection zone, leads to
\begin{eqnarray}
\RaFS = \frac{\Fbot}{8\rho \Omega^3 \Hp^3}.\label{equ:RaFS_simple}
\end{eqnarray}

\subsection{Diagnostics quantities}

The global Reynolds and SGS P\'eclet numbers describe the strength of
advection versus viscosity and SGS diffusion
\begin{eqnarray}
\Rey = \frac{\urms}{\nu k_1},\ \ \ \Pe_{\rm SGS} = \frac{\urms}{\chiSGS k_1},
\end{eqnarray}
where $\urms$ is the volume averaged rms-velocity, and where
$k_1=2\pi/d$ is an estimate of the largest eddies in the system. The
Reynolds and P\'eclet number based on the actual convective length
scale $\ell$ are given by
\begin{eqnarray}
\Rel = \frac{\urms\ell}{\nu},\ \ \ \Pel = \frac{\urms\ell}{\chiSGS}.\label{equ:RelPel}
\end{eqnarray}
Here $\ell = \kmean^{-1}$ is chosen, where $\kmean=\kmean(z)$ is the
mean wavenumber \citep[e.g.][]{CA06,SPD12}, and which is computed from
\begin{eqnarray}
  \kmean(z) = \frac{\int k E(k,z) dk}{\int E(k,z) dk},\label{equ:kint}
\end{eqnarray}
where $E(k,z)$ is the power spectrum of the velocity field with
$\uuu^2(z) = \int E(k,z)dk$.

\begin{figure}
  \includegraphics[width=\columnwidth]{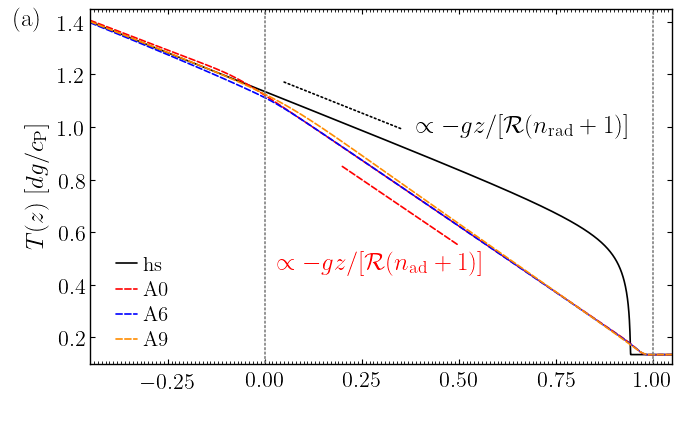}
  \includegraphics[width=\columnwidth]{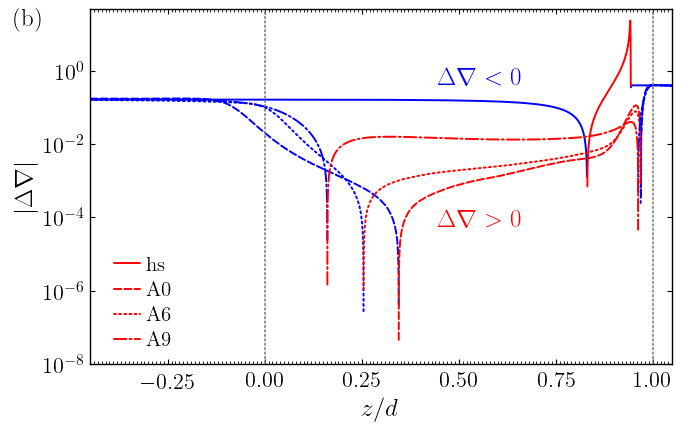}
\caption{\emph{(a)} Temperature as a function of height from a 1D
  hydrostatic model (black solid line) as well as convective runs
  Run~A0 (red dashed), A6 (blue dashed), and A9 (orange dashed). The
  black (red) dotted line shows a polytropic gradient corresponding to
  index $n_{\rm rad} = 3.25$ ($n_{\rm ad} = 1.5$) for
  reference. \emph{(b)} Absolute value of the superadiabatic
  temperature gradient $\Delta\nabla$ from the same runs as indicated
  by the legend. Red (blue) indicates convectively unstable (stable)
  stratification. The dotted vertical lines at $z=0$ and $z/d=1$
  denote the base and top of the initially isentropic layer.}
\label{fig:plot_hs}
\end{figure}

In general the total thermal diffusivity is given by
\begin{eqnarray}
  \chieff(\xxx) = \chiSGS + \chi(\xxx).
\end{eqnarray}
However, in all of the current simulations $\chi\ll\chiSGS$ in the CZ
such that the Prandtl and P\'eclet numbers based on $\chieff$ differ
very little from $\PraSGS$ and $\PeSGS$. The Rayleigh number is
defined as
\begin{eqnarray}
\Ra &=& \frac{gd^4}{\nu \chi}\left( - \frac{1}{\cP}\frac{{\rm d}s}{{\rm d}z} \right)_{\rm hs},
\end{eqnarray}
which varies as a function of height and is quoted near the surface at
$z/d=0.85$. The Rayleigh number in the hydrostatic, non-convecting,
state is measured from a one-dimensional model that is run to thermal
equilibrium, and where the convectively unstable layer is confined to
the near-surface layers \citep{Br16}; see also \Fig{fig:plot_hs}. In
the hydrostatic case $\chi=\chi(z)$ and $\chiSGS$, which affects only
the fluctuations, plays no role. The turbulent Rayleigh number is
quoted from the statistically stationary state using the horizontally
averaged mean state,
\begin{eqnarray}
\Rat &=& \frac{gd^4}{\nu \mchieff}\left.\left( - \frac{1}{\cP}\frac{{\rm
    d}\mean{s}}{{\rm d}z} \right)\right|_{z/d=0.85},
\end{eqnarray}
where the overbar denotes temporal and horizontal averaging.

Rotational influence on the flow is measured by several versions of
the Coriolis number. First, the global Coriolis number is defined as
\begin{eqnarray}
\Co = \frac{2 \Omega_0}{\urms k_1},\label{equ:Co}
\end{eqnarray}
where $k_1 = 2\pi/d$ is the wavenumber corresponding to the system
scale. This definition neglects the changing length scale as a
function of rotation and overestimates the rotational influence when
rotation is rapid and the convective scale is smaller. A definition
that takes the changing length scale into account is given by the
vorticity-based Coriolis number
\begin{eqnarray}
\Co_\omega = \frac{2 \Omega_0}{\orms},\label{equ:Cow}
\end{eqnarray}
where $\orms$ is the volume-averaged rms-value of the vorticity
$\bm\omega = \bm\nabla \times \uuu$. Another definition of the
Coriolis number taking into account the changing integral length scale
is given by
\begin{eqnarray}
\Col = \frac{2\Omega_0\ell}{\urms},\label{equ:Col}
\end{eqnarray}
where $\ell=(\mkmean)^{-1}$ where the overbar denotes averaging over
time and CZ. This is a commonly used choice in simulations of
convection in spherical shells \citep[][]{SPD12,GYMRW14}; see also
\cite{2020PhRvR...2d3115A} who considered convection in the limits of
slow rotation and rapid rotation.

Let us further define a flux Coriolis number $\CoF$\footnote{The same
  quantity was referred to as stellar Coriolis number in
  \cite{2023A&A...669A..98K}.} as
\begin{eqnarray}
\CoF \equiv \frac{2\Omega_0 \Hp}{\uflux} = 2\Omega_0 \Hp \left(\frac{\rhostar}{\Fbot}\right)^{1/3},\label{equ:CoF}
\end{eqnarray}
where $\uflux$ is a reference velocity obtained from
\begin{eqnarray}
\Ftot = \rho_\star \ustar^3,\label{equ:Ftot}
\end{eqnarray}
where $\rho_\star$ is a reference density, taken here at the bottom of
the CZ. $\uflux$ does not, and does not need to, correspond to any
actual velocity and it rather represents the available energy
flux. Therefore $\CoF$ does not depend on any dynamical flow speed
or length scale which are set by complicated interactions of
convection, rotation, magnetism, and other relevant physics. On the
other hand, $\CoF$ depends only on quantities that can either be
measured ($\Ftot$, $\Omega_0$) or deduced from stellar structure
models with relatively little ambiguity ($\Hp$, $\rho_\star$). The
significance of $\CoF$ is seen when rearranging
\Eq{equ:RaFS_simple} to yield
\begin{eqnarray}
(2\Omega_0 \Hp)^3\ \frac{\rho}{\Fbot} = (\RaFS)^{-1}.
\end{eqnarray}
Identifying the lhs with $\CoF^3$, \Eq{equ:CoF}, gives
\begin{eqnarray}
\CoF = (\RaFS)^{-1/3}.\label{equ:CoFRaFS}
\end{eqnarray}
An often used phrase in the context of convection simulations
targeting the Sun is that while all the other system parameters are
beyond the reach of current simulations, the rotational influence on
the flow can be reproduced
\citep[e.g.][]{2023SSRv..219...58K}. \Equ{equ:CoFRaFS} gives this a
more precise meaning in that the solar value of $\RaFS$ needs to be
matched by any simulation claiming to model the Sun.

The net vertical energy flux consists of contributions due to
radiative diffusion, enthalpy, kinetic energy flux, and viscous fluxes
as well as the surface cooling:
\begin{eqnarray}
\mFrad  &=& - \mean{K} \frac{{\rm d} \mean{T}}{{\rm d} z},\label{equ:mFrad}\\
\mFenth &=& \cP \mean{(\rho u_z)' T'},\label{equ:mFenth}\\
\mFkin  &=& \onehalf \mean{\rho \uuu^2 u_z'},\label{equ:mFkin}\\
\mFvisc &=& -2 \nu \mean{\rho u_i \mathsf{S}_{iz}}\label{equ:mFvisc}\\
\mFcool &=& - \int_{z_{\rm bot}}^{z_{\rm top}} \mean{\mathcal{C}} {\rm d}z.\label{equ:mFcool}
\end{eqnarray}
Here the primes denote fluctuations and overbars horizontal
averages. The total convected flux \citep{CBTMH91} is the sum of the
enthalpy and kinetic energy fluxes:
\begin{eqnarray}
\mFconv = \mFenth + \mFkin,
\end{eqnarray}
which corresponds to the convective flux in, for example, mixing
length models of convection.

Another useful diagnostic is buoyancy or Brunt-V\"ais\"al\"a
frequency, which is given by
\begin{eqnarray}
N^2 = \frac{g}{\cP}\frac{ds}{dz},
\end{eqnarray}
and describes the stability of an atmosphere with respect to buoyancy
fluctuations if $N^2>0$. Finally, the Richardson number related to
rotation in the stably stratified layers is defined as
\begin{eqnarray}
  \Ri_\Omega = \frac{N^2}{\Omega_0^2}.
\end{eqnarray}
Averages denoted by overbars are typically taken over the horizontal
directions and time, unless specifically stated otherwise.

\begin{table*}[t!]
\centering
\caption[]{Summary of the runs.}
  \label{tab:runs1}
      $$
          \begin{array}{p{0.05\linewidth}ccccccccccccccc}
          \hline
          \hline
          \noalign{\smallskip}
Run  & \RaF [10^{13}]  & \RaFS  & \Ta  & \Fn[10^{-6}] &  \Co  & \Cow  & \Co_\ell  & \CoF  & \Rey  & \Rat [10^6]  & \Ri_\Omega^{\rm RZ}  & \mbox{grid} \\
\hline
A0      &  0.5  &         -           &        0         & 4.6 &  0.00  &   0.00  &   0.00  &   0.00  &   38.7  &    4.1  &        -         &  288^3 \\
A1      &  0.5  &  2.5 \cdot 10^2     &  1.0 \cdot 10^4  & 4.6 &  0.07  &   0.02  &   0.08  &   0.28  &   38.9  &    4.3  &  1.4 \cdot 10^4  &  288^3 \\
A2      &  0.5  &         31          &  4.0 \cdot 10^4  & 4.6 &  0.13  &   0.03  &   0.16  &   0.55  &   39.6  &    4.3  &  3.3 \cdot 10^3  &  288^3 \\
A3      &  0.5  &         3.9         &  1.6 \cdot 10^5  & 4.6 &  0.26  &   0.06  &   0.31  &   1.04  &   39.4  &    4.6  &  8.0 \cdot 10^2  &  288^3 \\
A4      &  0.5  &        0.25         &  1.0 \cdot 10^6  & 4.6 &  0.63  &   0.15  &   0.71  &   2.38  &   40.0  &    5.1  &  1.2 \cdot 10^2  &  288^3 \\
A5      &  0.5  &        0.11         &  1.7 \cdot 10^6  & 4.6 &  0.82  &   0.19  &   0.87  &   3.04  &   40.0  &    5.5  &  7.1 \cdot 10^1  &  288^3 \\
A6      &  0.5  &  3.1 \cdot 10^{-2}  &  4.0 \cdot 10^6  & 4.6 &  1.27  &   0.29  &   1.26  &   4.56  &   39.8  &    6.1  &         30       &  288^3 \\
A7      &  0.6  &  3.9 \cdot 10^{-3}  &  1.6 \cdot 10^7  & 4.6 &  2.62  &   0.57  &   2.17  &   9.17  &   38.7  &    7.5  &        7.5       &  288^3 \\
A8      &  0.6  &  2.5 \cdot 10^{-4}  &  1.0 \cdot 10^8  & 4.6 &  7.21  &   1.38  &   3.88  &   25.1  &   35.1  &   10.9  &        1.2       &  288^3 \\
A9      &  0.7  &  3.0 \cdot 10^{-5}  &  4.0 \cdot 10^8  & 4.6 &  16.5  &   2.67  &   6.31  &   52.9  &   30.7  &   16.0  &        0.32      &  288^3 \\
\hline
B0      &  1.9  &         -           &        0         & 1.8 &  0.00  &   0.00  &   0.00  &   0.00  &   38.3  &    4.4  &        -         &  288^3\\
B1      &  1.9  &  2.5 \cdot 10^2     &  1.0 \cdot 10^4  & 1.8 &  0.07  &   0.02  &   0.08  &   0.27  &   38.5  &    4.4  &  2.5 \cdot 10^4  &  288^3 \\
B2      &  1.9  &         32          &  4.0 \cdot 10^4  & 1.8 &  0.13  &   0.03  &   0.17  &   0.54  &   38.9  &    4.5  &  6.2 \cdot 10^3  &  288^3 \\
B3      &  1.8  &         4.0         &  1.6 \cdot 10^5  & 1.8 &  0.26  &   0.06  &   0.32  &   1.03  &   39.0  &    4.8  &  1.5 \cdot 10^3  &  288^3 \\
B4      &  1.8  &        0.26         &  1.0 \cdot 10^6  & 1.8 &  0.65  &   0.15  &   0.70  &   2.34  &   39.3  &    5.3  &  2.2 \cdot 10^2  &  288^3 \\
B5      &  1.8  &  9.3 \cdot 10^{-2}  &  2.0 \cdot 10^6  & 1.8 &  0.90  &   0.20  &   0.97  &   3.21  &   39.4  &    5.7  &  1.1 \cdot 10^2  &  288^3 \\
B6      &  1.8  &  3.2 \cdot 10^{-2}  &  4.0 \cdot 10^6  & 1.8 &  1.30  &   0.29  &   1.27  &   4.49  &   39.1  &    6.3  &         55       &  288^3 \\
B7      &  1.9  &  4.0 \cdot 10^{-3}  &  1.6 \cdot 10^7  & 1.8 &  2.69  &   0.56  &   2.09  &   8.98  &   37.7  &    7.8  &         14       &  288^3 \\
B8      &  2.0  &  2.6 \cdot 10^{-4}  &  1.0 \cdot 10^8  & 1.8 &  7.32  &   1.36  &   3.90  &   24.4  &   34.6  &   11.4  &        2.3       &  288^3 \\
B9      &  2.2  &  3.1 \cdot 10^{-5}  &  4.0 \cdot 10^8  & 1.8 &  16.6  &   2.67  &   6.32  &   51.4  &   30.5  &   16.2  &        0.58      &  288^3 \\
\hline
C0      &  5.0  &         -           &        0         & 0.9 &  0.00  &   0.00  &   0.00  &   0.00  &   38.1  &    4.7  &        -         &  288^3 \\
C1      &  5.0  &  2.6 \cdot 10^2     &  1.0 \cdot 10^4  & 0.9 &  0.07  &   0.01  &   0.08  &   0.27  &   38.2  &    4.8  &  4.0 \cdot 10^4  &  288^3 \\
C2      &  4.9  &         32          &  4.0 \cdot 10^4  & 0.9 &  0.13  &   0.03  &   0.16  &   0.53  &   38.4  &    4.9  &  9.8 \cdot 10^3  &  288^3 \\
C3      &  4.8  &         4.0         &  1.6 \cdot 10^5  & 0.9 &  0.26  &   0.06  &   0.31  &   1.01  &   38.7  &    5.0  &  2.3 \cdot 10^3  &  288^3 \\
C4      &  4.7  &        0.26         &  1.0 \cdot 10^6  & 0.9 &  0.65  &   0.14  &   0.73  &   2.31  &   39.2  &    5.5  &  3.5 \cdot 10^2  &  288^3 \\
C5      &  4.7  &  1.2 \cdot 10^{-1}  &  1.7 \cdot 10^6  & 0.9 &  0.84  &   0.19  &   0.91  &   2.96  &   39.1  &    5.7  &  2.1 \cdot 10^2  &  288^3 \\
C6      &  4.7  &  3.2 \cdot 10^{-2}  &  4.0 \cdot 10^6  & 0.9 &  1.31  &   0.28  &   1.27  &   4.47  &   38.7  &    6.3  &         87       &  288^3 \\
C7      &  4.8  &  4.1 \cdot 10^{-3}  &  1.6 \cdot 10^7  & 0.9 &  2.71  &   0.55  &   2.14  &   8.93  &   37.3  &    8.1  &         22       &  288^3 \\
C8      &  5.0  &  2.6 \cdot 10^{-4}  &  1.0 \cdot 10^8  & 0.9 &  7.43  &   1.35  &   4.03  &   23.8  &   34.1  &   11.6  &        3.5       &  288^3 \\
C9      &  5.3  &  3.2 \cdot 10^{-5}  &  4.0 \cdot 10^8  & 0.9 &  16.8  &   2.65  &   6.44  &   50.5  &   30.1  &   16.7  &        0.91      &  288^3 \\
\hline
A1m     &  1.1  &  2.5 \cdot 10^2     &  4.0 \cdot 10^4  & 4.6 &   0.06  &   0.01  &   0.07  &   0.29  &   83.2  &   20.6  &  1.4 \cdot 10^4  &  576^3 \\
A3m     &  1.0  &         3.9         &  6.4 \cdot 10^5  & 4.6 &  0.24  &   0.04  &   0.27  &   1.05  &   84.3  &   21.1  &  8.0 \cdot 10^2  &  576^3 \\
A5m     &  1.1  &        0.12         &  6.8 \cdot 10^6  & 4.6 &  0.79  &   0.14  &   0.78  &   3.00  &   83.3  &   23.1  &        71        &  576^3 \\
A6m     &  1.1  &  3.2 \cdot 10^{-2}  &  1.6 \cdot 10^7  & 4.6 &  1.21  &   0.21  &   1.08  &   4.53  &   83.5  &   24.8  &        30        &  576^3 \\
A7m     &  1.1  &  4.0 \cdot 10^{-3}  &  6.4 \cdot 10^7  & 4.6 &  2.50  &   0.40  &   1.88  &   9.06  &   81.1  &   31.6  &       7.5        &  576^3 \\
A8m     &  1.2  &  2.5 \cdot 10^{-4}  &  4.0 \cdot 10^8  & 4.6 &  6.83  &   0.99  &   3.51  &   24.8  &   74.2  &   45.4  &       1.2        &  576^3 \\
A9m     &  1.3  &  3.1 \cdot 10^{-5}  &  1.6 \cdot 10^9  & 4.6 &  15.5  &   1.96  &   5.59  &   53.6  &   65.6  &   59.5  &       0.31       &  576^3 \\
\hline
A5h     &  2.1  &        0.11         &  2.7 \cdot 10^7  & 4.6 &  0.76  &   0.10  &   0.68  &   3.00  &    174  &   91.4  &        71        & 1152^3 \\
A9h     &  2.5  &  3.0 \cdot 10^{-5}  &  6.4 \cdot 10^9  & 4.6 &  14.2  &   1.43  &   8.36  &   54.5  &    143  &    245  &       0.31       & 1152^3 \\
\hline
          \end{array}
          $$ \tablefoot{Summary of the runs. $\PraSGS =1$ in all runs
            such that $\PeSGS=\Rey$.  Runs with rotation were made
            with $\theta = 0$, corresponding to the north pole of the
            star. $\taucool=2.5(\Fn/\Fn^{\rm A})\sqrt{d/g}$, where
            $\Fn^{\rm A}=4.6\cdot 10^{-6}$ is the normalized flux in
            the runs in Set~A.}
\end{table*}

\section{Results} \label{sect:results}

Three sets of simulations with varying $\Fn$ and approximately the
same values of $\CoF$ are presented. These will be referred to as
Sets~A, B, and C. The non-rotating runs in these sets correspond
respectively to Runs~K3, K4, and K5 in \cite{2019A&A...631A.122K} in
terms of $\Fn$, although lower values of $\nu$ and $\chiSGS$ were used
in the runs of the present study. Note that when $\Fn$ is varied
between the sets of simulations, the rotation rate $\Omega_0$, and the
diffusivities $\nu$ and $\chiSGS$ are varied at the same time
proportional to $\Fn^{1/3}$ \citep[see, e.g.][and
  \Appendix{app:convsca} for more
  details]{2020GApFD.114....8K}. Furthermore, the cooling time
$\taucool$ is varied proportional to $\Fn$. The current simulations
have modest Reynolds and P\'eclet numbers in comparison to
astrophysically relevant parameter regimes
\citep[e.g.][]{O03,2017LRCA....3....1K,2023SSRv..219...58K}; see
\Table{tab:runs1}. Earlier studies from non-rotating convection
suggest that results obtained at such modestly turbulent regimes
remain robust also at the highest resolutions affordable
\citep{2021A&A...655A..78K}. This is due to the fact that the main
energy transport mechanism (convection) and the main driver of
convection (surface cooling) are not directly coupled to the
diffusivities. However, the current cases with rotation are more
complicated because the supercriticality of convection decreases with
increasing rotation rate \citep[e.g.][]{Ch61,1968RSPTA.263...93R}.
The effects of decreasing supercriticality are not studied
systematically here, but subsets of the runs in Set A were repeated
with higher resolutions ($576^3$ and $1152^3$) and correspondingly
higher $\RaF$, $\Rey$, and $\Pe$; see Sets~Am and Ah in
\Table{tab:runs1}.

\begin{figure*}
  \includegraphics[width=\textwidth]{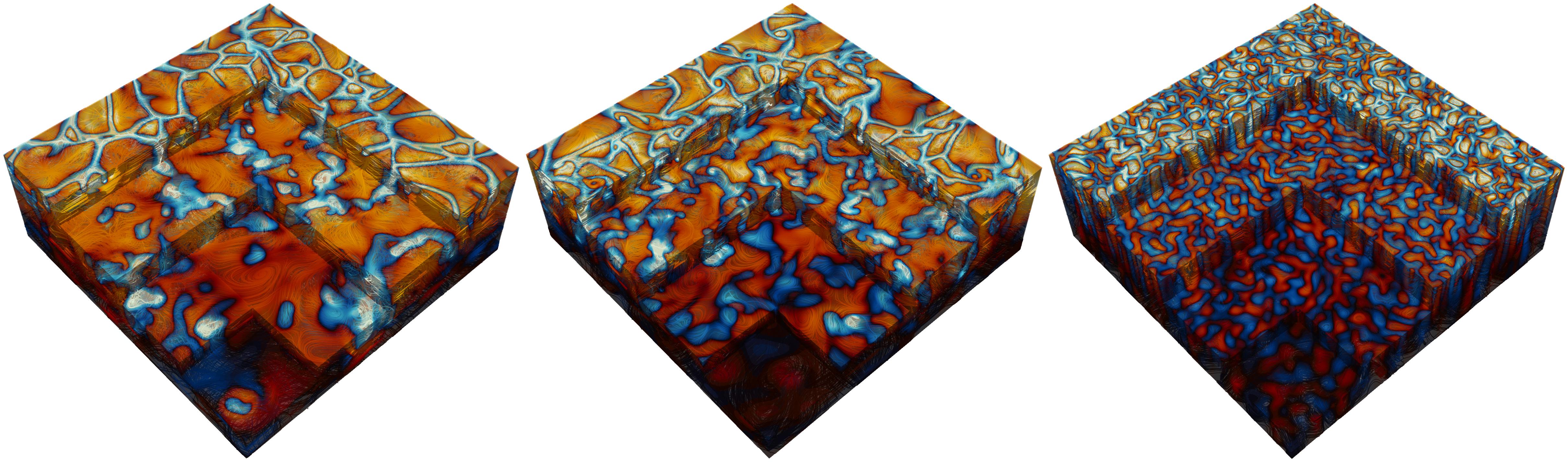}
  \caption{Flow fields from Runs~A2 with $\Co = 0.13$ (left), A6 with
    $\Co = 1.3$ (middle), and A9 $\Co = 16.5$ (right) at the north
    pole ($\theta = 0\degr$). The colours indicate vertical velocity
    and the contours indicate streamlines.}
\label{fig:boxes}
\end{figure*}

\subsection{Hydrostatic solution}

Earlier studies have shown that a purely radiative hydrostatic
solution with the Kramers opacity law is a polytrope with index
$n_{\rm rad}=3.25$ \citep{BB14,Br16}. Such a solution arises in the
case where $K = {\rm const.}$ and $\nabla_z T = {\rm const.}$ To see
if this configuration is recovered with the current set-up,
\Equs{equ:dens}{equ:ent} were solved numerically in a one-dimensional
$z$-dependent model with otherwise the same parameters as in the 3D
simulations corresponding to the runs in Set~A. The resulting
temperature profile is shown in \Figa{fig:plot_hs}(a) along with a
corresponding horizontally averaged profile from convecting Runs~A0,
A6, and A9. The stratification is consistent with a polytrope
corresponding to $\nrad$ up to a height of roughly $z/d=0.75$. Near
the nominal surface of the convection zone, $z/d=1$, the temperature
gradient steepens sharply because the cooling term relaxes the
temperature toward a constant ($z$-independent) value near the
surface. Therefore, neither $K$ nor $\nabla_z T$ are constants in this
transition region between the radiative and the cooling layers. In the
initial state the stratification is isothermal above $z/d =1$, but
because the cooling profile $f_{\rm cool}$ has a finite width, cooling
also occurs below $z/d =1$ and the isothermal layer is wider in the
final thermally saturated state. This also depends on the value of
$\taucool$. In the convective runs the stratification is nearly
polytropic with index $\nrad$ near the base of the radiative layer and
nearly isentropic with $\nad$ in the bulk of the convection zone.

The superadiabatic temperature gradient is defined as
\begin{eqnarray}
  \Delta\nabla \equiv \nabla - \nabad = -\frac{\Hp}{\cP}\frac{{\rm d}s}{{\rm d}z},
\end{eqnarray}
where $\nabla = {\rm d} \ln \mean{T}/{\rm d} \ln \mean{p}$ is the
logarithmic temperature gradient and $\nabad = 1 - 1/\gamma$ is the
corresponding adiabatic gradient. Comparison of the hydrostatic
profile and the non-rotating convective model A0 shows that the
convectively unstable layer in the former is much thinner than in the
latter. This is a direct consequence of the strong temperature and
density dependence of the Kramers opacity law. A similar conclusion
applies also to the Sun, where the hypothetical non-convecting
hydrostatic equilibrium solution has a very thin superadiabatic layer
\citep{Br16}. The steepness of the temperature gradient near the
surface is characterized by the maximum value of $(\Delta\nabla)_{\rm
  hyd}$, which is 23.4. By comparison, in the convective Run~A0
$\Delta\nabla = 0.12$. The Rayleigh number -- measured at $z/d=0.85$
-- in the hydrostatic case is $\Ra = 5.4\cdot10^7$, which is about an
order of magnitude greater than $\Rat$ in Run~A0.

\subsection{Qualitative flow characteristic as a function of rotation}

\Figu{fig:boxes} shows representative flow fields from runs with slow,
intermediate, and rapid rotation, corresponding to Coriolis numbers
$0.13$, $1.3$, and $16.5$, respectively. The effects of rotation are
hardly discernible in the slowly rotating case A2 with $\Co=0.13$. In
the run with intermediate rotation, Run~A6 with $\Co=1.3$, the
convection cells are somewhat smaller than in the slowly rotating case
and more vortical structures are visible near the surface. For the
most rapidly rotating case, Run~A9 with $\Co=16.5$, the size of the
convection cells is drastically reduced in comparison to the other two
runs and clear alignment of the convection cells with the rotation
vector is seen.

\begin{figure*}
  \includegraphics[width=\textwidth]{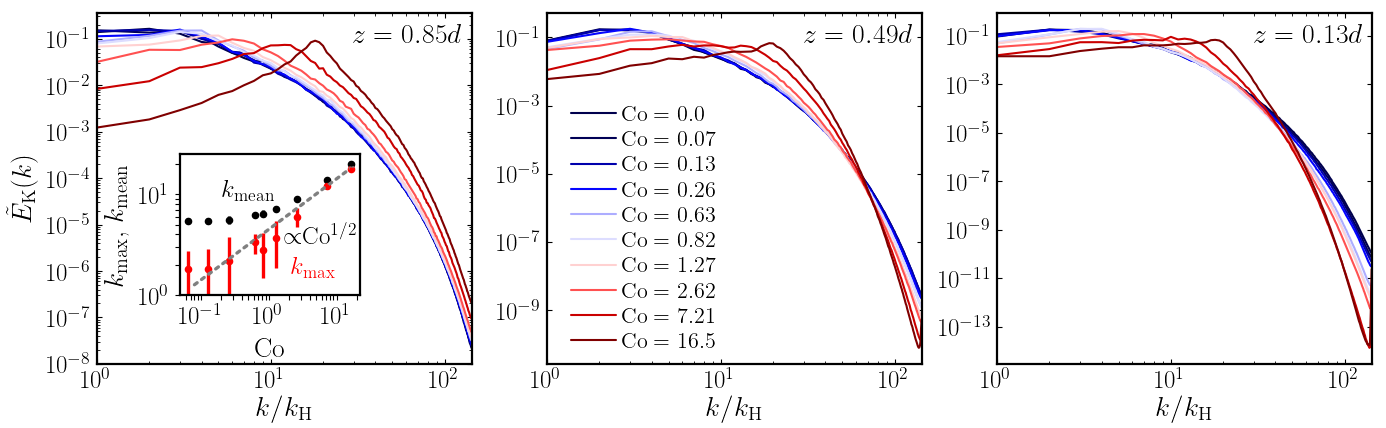}
\caption{Normalized velocity power spectra near the surface (left
  panel), middle (middle), and base (right) of the CZ from runs in Set
  A with $\Co$ varying between $0$ and $16.5$. The inset in the left
  panel shows the mean scale $\kmean$ and wavenumber of the where
  $E(k)$ has its maximum ($\kmax$) as functions of $\Co$ for $z/d
  =0.85$. The error bars indicate the standard deviation. The gray
  dashed line shows a power law proportional to $\Co^{1/2}$.}
\label{fig:plot_uu_spectra}
\end{figure*}

\subsection{Convective scale as function of rotation}

Power spectra $E(k)$ of the velocity fields for the runs in Set~A are
shown in \Figa{fig:plot_uu_spectra} from depths near the surface, at
the middle and near the base of the CZ. As was already evident from
visual inspection of the flow fields, the dominant scale of the flow
decreases as the rotation rate increases. Quantitatively, the
wavenumber $\kmax$, where $E(k)$ has its maximum, increases roughly in
proportion to $\Co^{1/2}$. The mean wavenumber $\kmean$, computed
from \Eq{equ:kint}, shows the same scaling for $\Co \gtrsim 2$. This
is explained by the broader distribution of power at different
wavenumbers at slow rotation in comparison to the rapid rotation cases
where fewer -- or just a single -- convective modes are dominant; see
\Fig{fig:plot_uu_spectra}.

A decreasing length scale of the onset of linear instability under the
influence of rotation was derived in \citep{Ch61} with $k_{\rm onset}
\propto \Ta^{2/3}$. With $\Ta \propto \Co^2 \Rey^2$, and with $\Rey$
being approximately constant, $k_{\rm onset} \propto \Co^{1/3}$ is
obtained. On the other hand, considering the CI part of the CIA
balance in the Navier--Stokes equation gives \citep[e.g.][see
  \Eq{equ:kmaxk1} in \Appendix{app:convsca}]{2020PhRvR...2d3115A}
gives
\begin{eqnarray}
\left(\frac{\kmax}{k_1}\right)^2 \propto \frac{2\Omega}{k_1 u} = \Co,
\end{eqnarray}
or $\kmax \propto \Co^{1/2}$. This is consistent with the current
simulations; see the inset in the left panel of
\Figa{fig:plot_uu_spectra}. The same result was obtained in
\cite{FH16b}. Some nonlinear convection simulations show scalings that
are similar but somewhat shallower than that obtained from the CIA
balance; see, e.g., \cite{2018A&A...616A.160V} and
\cite{2020MNRAS.493.5233C}.

To estimate the convective length scale in the Sun based on the
current results requires that the value of $\CoF$ matches that of the
deep solar CZ. The quantities on the rhs of \Eq{equ:CoF} at the base
of the solar convection zone are $\Hp \approx 5\cdot 10^7$~m,
$\rho_\star \approx 200$~kg~m$^{-3}$, $\Fbot = \Lsun/(4\pi r_{\rm
  CZ}^2) \approx 1.27\cdot 10^8$~kg~s$^{-3}$, with $r_{\rm CZ} =
0.7\Rsun \approx 4.9\cdot 10^{8}$~m and $\Lsun = 3.83\cdot 10^{26}$~W,
and $\Omsun = 2.7 \cdot 10^{-6}$~s$^{-1}$. Inserting this data into
\Eq{equ:CoF} yields $\CoF^{\odot} \approx 3.1$. The values of $\CoF$
are listed for all runs in the eight column of \Table{tab:runs1}. The
moderately rotating runs [A,B,C]5 correspond to the rotational
constraint at the base of the solar CZ with $\CoF=3.0\ldots3.2$. The
mean wavenumber $\kmean/k_1 \approx 7$ in these simulations
corresponds to a horizontal scale of $\ell_{\rm conv} = L_x
(k_1/\kmean) \approx 0.57d$. The pressure scale height at $z_{\rm DZ}$
is about $0.49d$ such that $\ell_{\rm conv} \approx
1.16\Hp$. Converting this to physical units using $\Hp^\odot \approx
5\cdot10^6$~m yields $\ell_{\rm conv} \approx 58$~Mm. Following the
procedure of \cite{FH16b} and using $\kmax$ instead of $\kmean$,
$\kmax/k_1=3$ and $\ell_{\rm conv} \approx 130$~Mm. Both of these
estimates are significantly larger that the supergranular scale of
$20\ldots 30$~Mm which was suggested to be the largest convectively
driven scale in the Sun by \cite{FH16b} and
\cite{2021PNAS..11822518V}. On the other hand, a rapidly rotating
run of \cite{FH16b} with $\lconv\approx30$~Mm, had Rossby number
$\Ro_{\rm FH} = \tilde{U}/(2\Omega H) = 0.011$, where $\tilde{U}$ is a
typical velocity amplitude and $H$ is the shell thickness. This
corresponds to a global Coriolis number $\Co = 2\pi \Ro_{\rm FH}^{-1}
\approx 14.5$ in the conventions of the current study. In the current
runs A9, B9, and C9, $\Co\approx17$ and $\kmax\approx\kmean\approx17$,
corresponding to $\lconv\approx26$~Mm. Therefore the current
simulations give a very similar estimate for $\lconv$ at comparable
values of $\Co$ despite all of the differences between the model
set-ups. However, the values of $\CoF$ in runs A9, B9, and C9 are at
least 16 times higher than in the Sun, suggesting that the simulations
of \cite{FH16b} were also rotating much faster than the
Sun\footnote{For example, their run with $\Ro = 0.011$ has
  $\RaF=6.81\cdot 10^6$, $\Ek=1.91\cdot10^{-4}$, and $\Pra=1$, and
  corresponds to $\RaFS=\RaF\Ek^3/(8\Pra)=5.9\cdot10^{-6}$, or $\CoF =
  (\RaFS)^{-1/3}\approx 55$. This yields $\Omega/\Omsun \approx
  [(\RaFS)_\odot/\RaFS]^{1/3}\approx 17.6$, which is approximate
  because different length scales are used.}. Therefore the current
results suggest that rotationally constrained convection cannot
explain the appearance of supergranular scale as the largest
convective scale in the Sun.

\begin{figure}
  \includegraphics[width=\columnwidth]{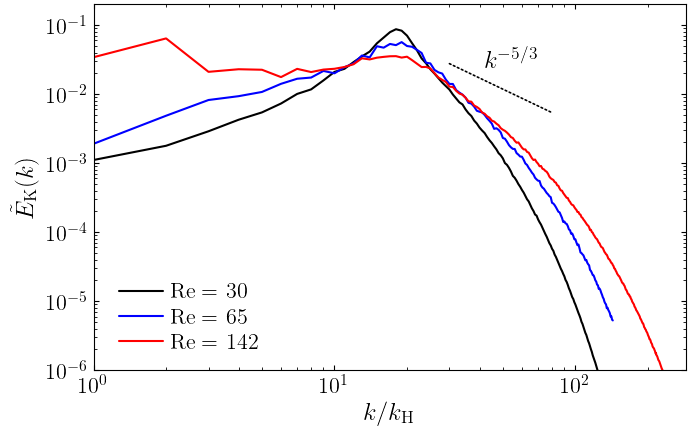}
\caption{Normalized velocity power spectra near the surface of
  simulations with $\Co\approx17$ and $\Rey=30\ldots 142$ (Runs~A9,
  A9m, and A9h). The dotted line shows a Kolmogorov $k^{-5/3}$ scaling
  for reference.}
\label{fig:plot_uu_spectra_Om1}
\end{figure}

\Figu{fig:plot_uu_spectra_Om1} shows the velocity power spectra for
the most rapidly rotating runs with $\Co\approx17$ for
$\Rey=30\ldots142$ from Runs~A9, A9m, and A9h. There is a marked
increase in the power at large scales, which begins to affect $\kmean$
at the highest $\Rey$ or Run~A9h. This is due to the gradual onset of
large-scale vorticity production, most likely due to
two-dimensionalisation of turbulence, that has been observed in
various earlier studies of rapidly rotating convection
\citep[e.g.][]{Chan03,Chan07,2013E&PSL.371..212C,2011ApJ...742...34K,GHJ14}. Despite
the rapid rotation with Coriolis numbers exceeding $16$, the
large-scale vorticity generated in the current simulations is
relatively modest apart from Run~A9h. A difference to many of the
previous studies is that here the relevant thermal Prandtl number
($\PraSGS$) is of the order of unity whereas in many of the earlier
studies $\Pra$ was lower. Large-scale vorticity production was indeed
observed in an additional run which is otherwise identical to A9
except that $\PraSGS=0.2$ instead of $\PraSGS=1$ (not shown).

\subsection{Measures of rotational influence}

\subsubsection{Velocity-based $\Co$}

\begin{figure}
  \includegraphics[width=\columnwidth]{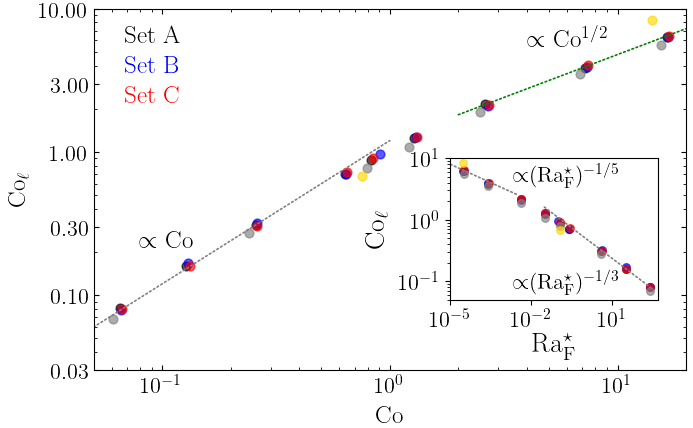}
\caption{Dynamical Coriolis number $\Col$ as a function of $\Co$ for
  all of the runs in \Table{tab:runs1}. Power laws proportional to
  $\Co$ (slow rotation; $\Co<1$) and $\Co^{1/2}$ (rapid rotation;
  $\Co>2$) are shown for reference. The inset shows $\Col$ as a
  function of $\RaFS$ with power laws proportional to $(\RaFS)^{-1/5}$
  (rapid rotation; $\RaFS < 3\cdot10^{-3}$), and $(\RaFS)^{-1/3}$
  (slow rotation; $\RaFS > 0.03$).}
\label{fig:plot_cocol}
\end{figure}

The suitability of different measures of rotational influence on the
flow has been discussed in various works in the literature
\citep[e.g.][]{2023A&A...669A..98K}. A common -- and justified --
critique regarding the Coriolis number as defined in \Equ{equ:Co} is
that it does not appreciate the fact that $\lconv = \lconv(\Omega)$
\citep[e.g.][]{2021PNAS..11822518V}. The most straightforward way is
to measure the mean wavenumber and use
\Eq{equ:Col}. \Fig{fig:plot_cocol} shows $\Col$ as a function of $\Co$
for all run listed in \Table{tab:runs1}. For slow rotation, $\Co
\lesssim 1$, $\Col \propto \Co$ because $\uconv$ and $\lconv$ are
almost unaffected by rotation. For sufficiently rapid rotation this is
no longer true because $\kmean\approx\kmax\propto \Co^{1/2}$ as
indicated by \Eq{equ:kmaxk1} and the simulation results; see the inset
of \Fig{fig:plot_uu_spectra}. This implies that for rapid rotation
$\Col \propto \Co^{1/2}$; see also \Eq{equ:colco}. This is consistent
with the numerical results found in the most rapidly rotating cases;
see \Figa{fig:plot_cocol}. The higher resolution runs in Set~Am have
somewhat lower $\Col$ than the corresponding runs in Set~A because the
convective velocities in the higher resolution cases are higher. This
shows that the simulations are not yet in an asymptotic regime where
the results are independent of the diffusivities. This is further
demonstrated by the high resolution runs of Set~Ah: Run~A5h follows
the trend set by Run~A5m. The Run~A9h with a significantly higher
$\Col$ than in Runs~A9 and A9m is explained by the increasing $\kmean$
due to the large-scale vorticity generation in that
case. \cite{2020PhRvR...2d3115A} showed that the dynamical Rossby
number is related to the diffusion-free modified flux Rayleigh number
$\RaFS$, with different powers for slow and rapid rotation. The
corresponding derivations for the Coriolis number $\Col$ are presented
in \Appendix{app:convsca}, and which show that $\Col = (\RaFS)^{-1/3}$
(slow rotation) and $\Col = (\RaFS)^{-1/5}$ (rapid rotation). Both
scalings are also supported by the simulation results; see the inset
of \Fig{fig:plot_cocol}.

\subsubsection{Vorticity-based $\Co$}

Another commonly-used definition, \Equ{equ:Cow}, is used to take the
changing length scale automatically into account. However, $\Cow$
comes with a caveat which has apparently not been discussed hitherto
in the astrophysical literature. This is demonstrated by considering a
set of rotating systems at asymptotically high $\Rey$ where $\urms$ is
independent of $\Rey$. The forcing is assumed fixed by a constant
energy flux through the system, and the asymptotic value of $\urms$
when $\Rey \rightarrow \infty$ as $u_\infty$. Furthermore, in this
regime the mean kinetic energy dissipation rate
\begin{eqnarray}
\mepsK = 2\nu \mean{\bm{\mathsf{S}}^2},
\end{eqnarray}
where the overbar denotes a suitably defined average, tends to a
constant value when normalized by mean length and corresponding
rms-velocity
\citep[e.g.][]{1984PhFl...27.1048S,2015AnRFM..47...95V}. This value is
denoted as $\epsilon_\infty$. In low-Mach number turbulence, which is
a good approximation of stellar interiors, as well as the current
simulations with $\Ma\sim \mathcal{O}(10^{-2})$,
\begin{eqnarray}
\mepsK = \nu \mean{\bm\omega^2} = \nu \orms^2.\label{equ:mepsK}
\end{eqnarray}
From the definition of system scale Reynolds number it follows that
\begin{eqnarray}
\Rey = \frac{u_\infty}{\nu k_1} \propto \nu^{-1},
\end{eqnarray}
and from \Eq{equ:mepsK} that
\begin{eqnarray}
\orms = \left(\frac{\mepsK}{\nu}\right)^{1/2} = \left(\frac{\epsilon_\infty}{\nu}\right)^{1/2} \propto \nu^{-1/2} \propto \Rey^{1/2}.\label{eq:orms}
\end{eqnarray}
Using \Eq{equ:Cow} it is found that
\begin{eqnarray}
\Cow \propto \Rey^{-1/2},\ \ \mbox{or}\ \ \Co \propto \Rey^{1/2}\Cow.
\end{eqnarray}
This means that $\Cow \rightarrow 0$ as $\Rey \rightarrow \infty$ at
constant $\Co$, while the dynamics at large (integral) scales are
unaffected. Therefore $\Cow$ underestimates the rotational influence
at the mean scale $\kmean$ which dominates the dynamics, as opposed to
\Eq{equ:Co} overestimating it.

\begin{figure}
  \includegraphics[width=\columnwidth]{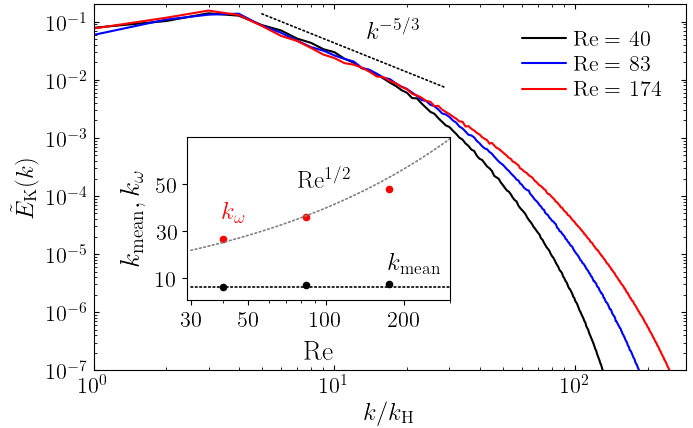}
\caption{Normalized velocity power spectra near the surface of
  simulations with $\Co\approx1.3$ and $\Rey=40\ldots 174$ (Runs~A6,
  A6b, and A6c). The dotted line shows Kolmogorov $k^{-5/3}$ scaling
  for reference. The inset shows $\kmean$ (black symbols) and
  $k_\omega$ (red) as functions of $\Rey$. The dotted lines are
  proportional to powers $0$, and $1/2$ of $\Rey$.}
\label{fig:plot_uu_spectra_Om0065}
\end{figure}

\Equ{eq:orms} can also be written as
\begin{eqnarray}
\orms \equiv k_\omega \urms \propto \Rey^{1/2}.\label{eq:komega}
\end{eqnarray}
For sufficiently large $\Rey$, the theoretical prediction is that
$\urms \rightarrow u_\infty = \mbox{const.}$ and $k_\omega \propto
\Rey^{1/2}$. This has been confirmed from numerical simulations of
isotropically forced homogeneous turbulence
\citep[e.g.][]{2012AN....333..195B,CB13}. Here the dependence of
$k_\omega$ on $\Rey$ is shown in the inset of
\Fig{fig:plot_uu_spectra_Om0065} for runs with $\Co\approx1.3$ and
$\Rey$ ranging between $40$ and $174$. Here the results for $k_\omega$
fall somewhat below theoretical $\Rey^{1/2}$ expectation. This is
likely because the asymptotic regime requires still higher Reynolds
numbers. On the other hand, the mean wavenumber $\kmean$ is
essentially constant around $\kmean/k_1 = 7$ in this range of $\Rey$
because the dominating contribution to the velocity spectrum come from
large scales that are almost unaffected by the increase in
$\Rey$.

\begin{figure}
  \includegraphics[width=\columnwidth]{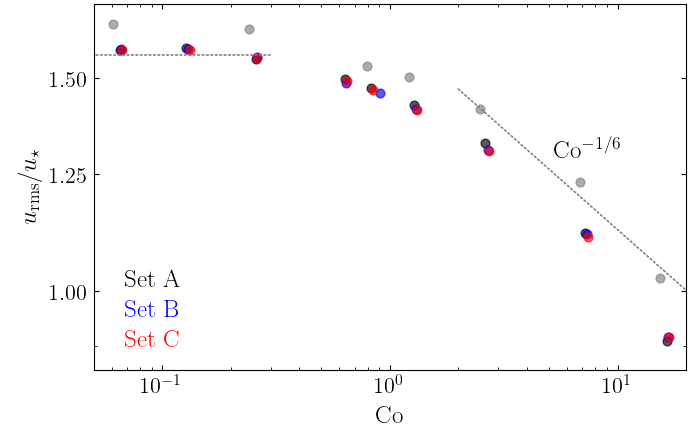}
\caption{Root-mean-square velocity in the convection zone normalized
  by $u_\star$. The dotted line is proportional to $\Co^{1/6}$ as
  indicated by the theoretical CIA scaling; see \Eq{equ:uFCo}.}
\label{fig:urms_Co}
\end{figure}

\begin{figure}
  \includegraphics[width=\columnwidth]{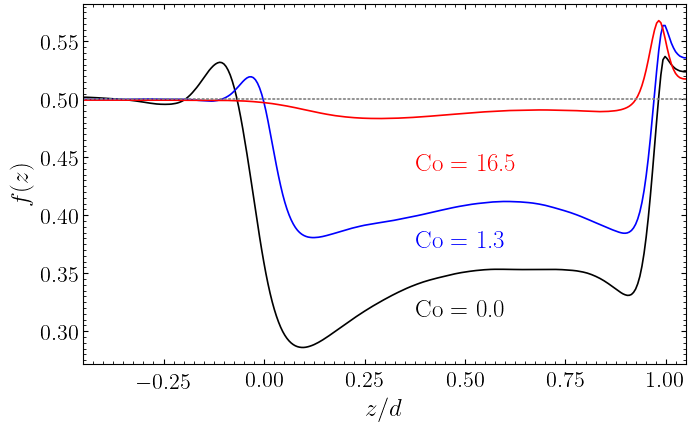}
\caption{Filling factor of downflows as a function of height $f(z)$
  for three runs with no (black), moderate (blue), and rapid (red)
  rotation.}
\label{fig:plot_ff}
\end{figure}

\subsection{Convective velocity as a function of total flux and rotation}

The scalings of convective velocity as a function of rotation are
derived in \Appendix{app:convsca} following the same arguments as in
\cite{2020PhRvR...2d3115A}. For slow rotation the convective velocity
depends only on the energy flux:
\begin{eqnarray}
\urms \sim \left(\frac{\Ftot}{\rho}\right)^{1/3} = \ustar,
\end{eqnarray}
where $\ustar$ is defined via \Eq{equ:Ftot}. This scaling is altered
in the rapidly rotating regime, where
\begin{eqnarray}
\urms \propto \left( \frac{F}{\rho} \right)^{1/3} \Co^{-1/6}.\label{equ:uFCo}
\end{eqnarray}
This results agrees with Eq.~(50d) \cite{2020PhRvR...2d3115A} and
Table~2 of \cite{2021PNAS..11822518V}. Therefore the velocity
amplitude in the rapidly rotating regime is expected to depend not
only on the available flux but also on rotation. \Figa{fig:urms_Co}
shows the corresponding numerical results for the Sets~A, B, C, and
Am. For slow rotation, $\Co \lesssim 0.3$, $\urms$ is roughly constant
around $\urms\approx 1.55\ustar$ for Sets~A, B, and C, and
$\urms\approx 1.65\ustar$ for Set~Am. In the rapid rotation regime
$\urms$ follows a trend which is similar to that indicated in
\Eq{equ:uFCo}, but the agreement is not perfect. The simulations in
this regime may suffer from the fact the supercriticality of
convection decreases with $\Co$. However, the medium resolution runs,
visualized by the grey symbols in \Figa{fig:urms_Co}, do not show a
significantly better agreement with theory. Nevertheless, the evidence
for CIA balance being reached in the current simulations with rapid
rotation is fairly convincing.

\begin{figure*}
  \includegraphics[width=\textwidth]{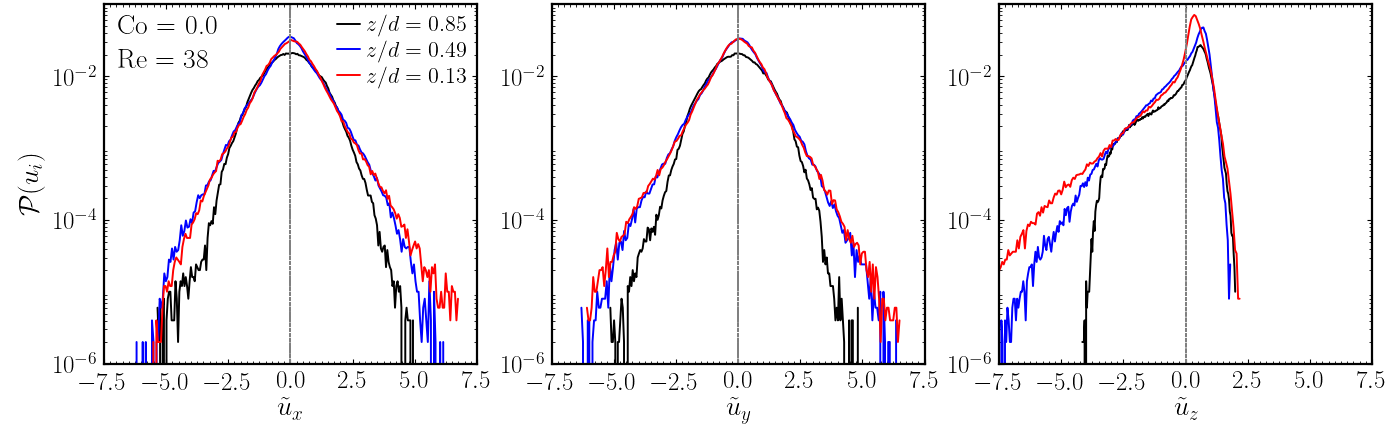}
  \includegraphics[width=\textwidth]{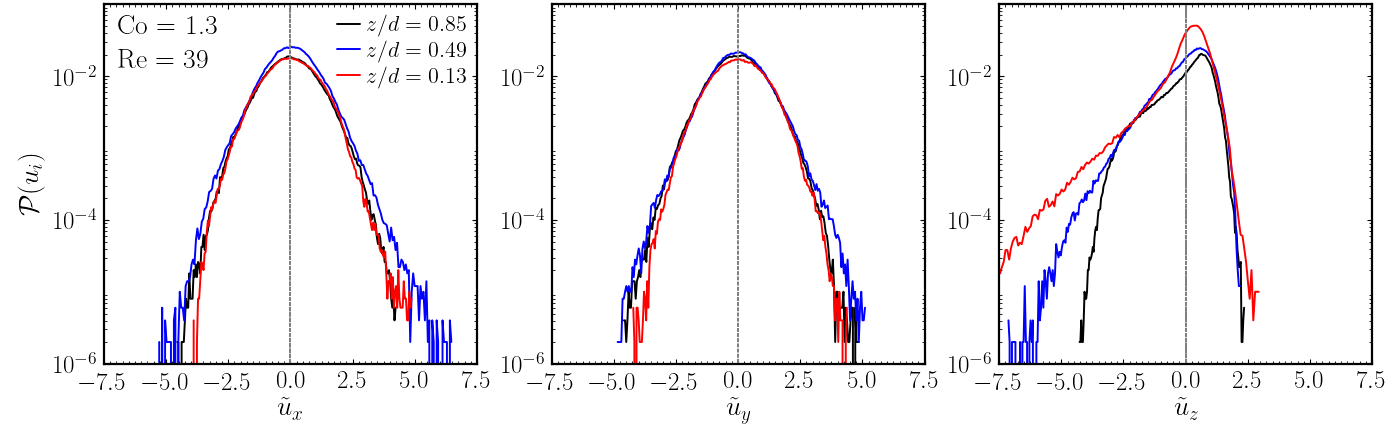}
  \includegraphics[width=\textwidth]{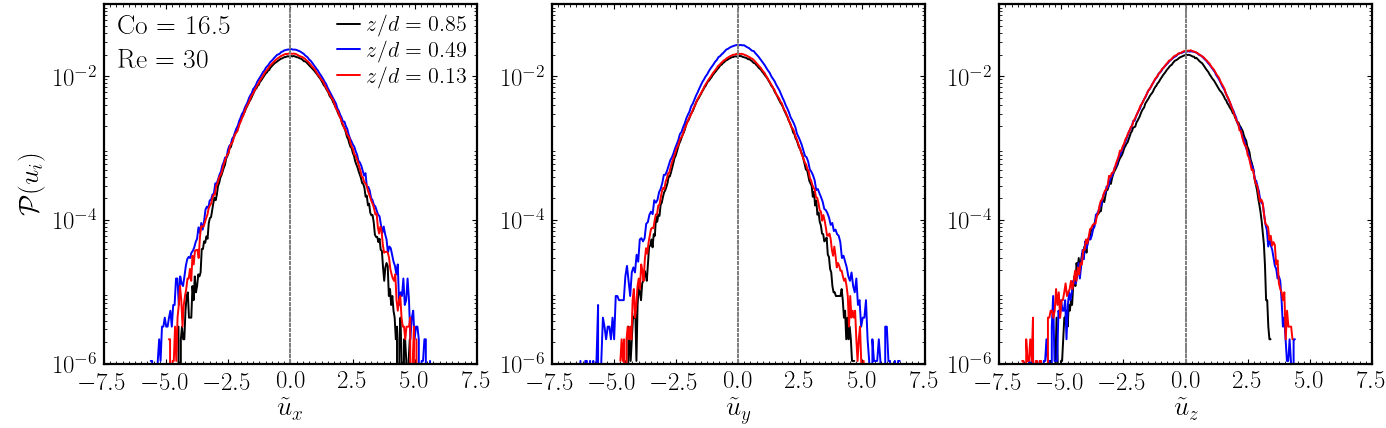}
\caption{Probability density functions $\mathcal{P}(u_i)$ for $u_x$
  (left), $u_y$ (middle), and $u_z$ (right) for depths $z/d = 0.85$
  (black), $z/d = 0.49$ (blue), and $z/d = 0.13$ (red) for runs with
  $\Co=0$ (Run~A0, top row), $\Co=1.3$ (Run~A6, middle), and $\Co =
  16.5$ (Run~A9, bottom). The tildes refer to normalization by the
  respective rms-values.}
\label{fig:plot_pdf}
\end{figure*}

\subsection{Flow statistics}

Compressible non-rotating convection is characterized by broad upflows
and narrow downflows \citep{SN89,CBTMH91}; see also
\Fig{fig:boxes}. This can be described by the filling factor $f$ of
downflows as
\begin{eqnarray}
\muuz(z) = f(z) \muuz^\downarrow + [1-f(z)] \muuz^\uparrow(z),
\end{eqnarray}
where $\muuz$ is the mean vertical velocity, whereas $\muuz^\uparrow$
and $\muuz^\downarrow$ are the corresponding mean up- and downflow
velocities. It was shown in \cite{2021A&A...655A..78K} that $f$ is
sensitive to the effective Prandtl number of the fluid such that a
lower $\Pra$ leads to a lower filling factor. Here a similar study is
done as a function of rotation; see \Figa{fig:plot_ff}. The main
result is that $f$ approaches $1/2$ in the rapid rotation regime. This
is because in rapidly rotating convection the broad upwellings of
non-rotating convection are broken up and the flow consist mostly of
smaller scale helical columns where the up- and downflows are almost
invariant. This is due to the Taylor-Proudmann constraint such that
derivatives along the rotation axis vanish. Hence the tendency for
larger structures to appear at greater depths is inhibited and the
average size of convection cells as a function of depth is almost
constant; see rightmost panel of \Fig{fig:boxes}.

\begin{figure*}
  \includegraphics[width=\textwidth]{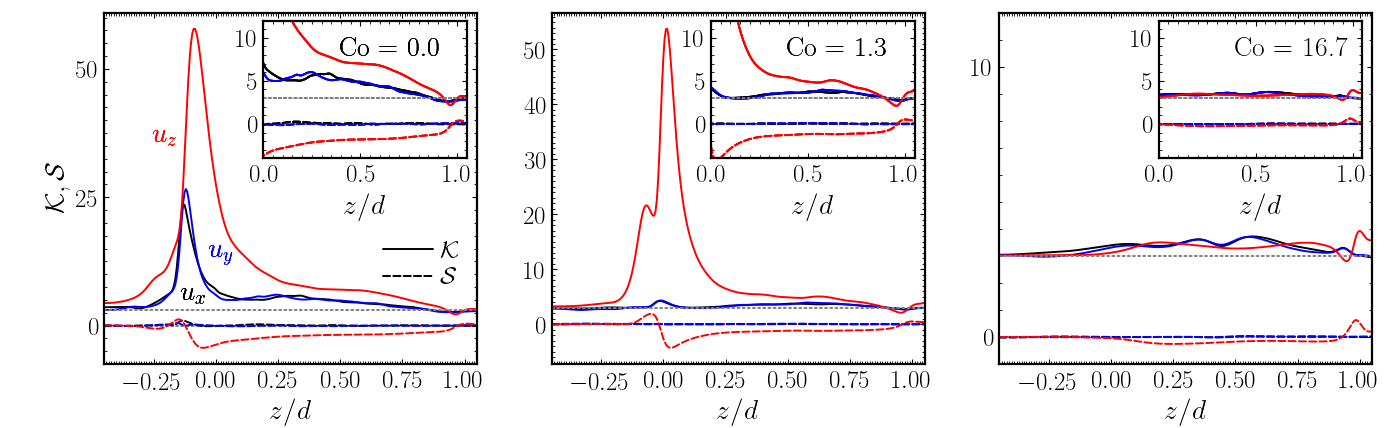}
\caption{Skewness ($\mathcal{S}$, dashed lines) and kurtosis
  ($\mathcal{K}$, solid) from the same runs as in
  \Fig{fig:plot_pdf}. Black, blue, and red colour indicates data
  corresponding to $u_x$, $u_y$, and $u_z$, respectively. Note the
  difference in scale between each of the panels. The insets show a
  zoom in of the region $z/d \ge 0$.}
\label{fig:plot_kurtskew}
\end{figure*}

This is also apparent from the probability density functions (PDFs) of
the velocity components $u_i$, defined via
\begin{eqnarray}
\int \mathcal{P}(u_i,z){\rm d}u_i = 1.
\end{eqnarray}
\Figu{fig:plot_pdf} shows representative examples of PDFs for the
extreme cases (Run~A0 with $\Co=0$ and Run~A9 with $\Co\approx16.5$)
and at an intermediate rotation rate (Run~A6, $\Co=1.3$). In
non-rotating convection the PDFs of the horizontal components of the
velocity are nearly Gaussian near the surface whereas for $u_z$ the
distributions are highly skewed due to the up-/downflow asymmetry. In
deeper parts also the horizontal velocities deviate from a Gaussian
distribution in agreement with earlier works
\citep[e.g.][]{BJNRST96,HRY15b,2021A&A...655A..78K}

As the rotation increases the asymmetry of the vertical velocity
decreases such that in the most rapidly rotating cases considered here
with $\Co \approx 17$, $u_z$ also approaches a Gaussian
distribution. Only near the surface ($z/d = 0.85$) a weak asymmetry
remains. The horizontal components of velocity continue to have
Gaussian distribution as rotation is increased, although there is not
enough data to say anything concrete concerning the tails of the
distributions at high velocity amplitudes. To further quantify the
statistics of the flow, skewness $\mathcal{S}$ and kurtosis
$\mathcal{K}$ are computed from:
\begin{eqnarray}
\mathcal{S} = \frac{\mathcal{M}^3}{\sigma_u^3},\ \mathcal{K} =
\frac{\mathcal{M}^4}{\sigma_u^4},
\end{eqnarray}
where $\sigma_u = (\mathcal{M}^2)^{1/2}$, with
\begin{eqnarray}
\mathcal{M}^n(u_i,z) = \int [u_i(\xxx) - \mean{u}_i(z)]^n \mathcal{P}(u_i,z){\rm d}u_i.
\end{eqnarray}
\Figu{fig:plot_kurtskew} shows $\mathcal{S}$ and $\mathcal{K}$ for all
$u_i$ for the same runs as in \Fig{fig:plot_pdf}. The skewness in
consistent with zero for the horizontal velocities which is expected
as there is not anisotropy in the horizontal plane. The negative
values of $\mathcal{S}$ for $u_z$ are a signature of the asymmetry
between up- and downflows. As rotation is increased, $\mathcal{S}$
approaches zero also for $u_z$. Kurtosis $\mathcal{K}$ is a measure of
non-Gaussianity or intermittency. In the non-rotating case
$\mathcal{K}$ increases from roughly three -- indicating Gaussian
statistics -- to roughly five for horizontal flows as a function of
depth within the CZ. For $u_z$ the increase of $\mathcal{K}$ is much
more dramatic below $z/d \lesssim 0.3$. This is because downflows
merge at deeper depths such that only a few of them survive deep in
the CZ and especially in the overshoot region below roughly $z=0$,
where $\mathcal{K}$ reaches a peak value of rouhgly 65 for Run~A0. A
similar, albeit lower, maximum appears also for the horizontal
flows. At intermediate rotation (Run~A6; $\Co=1.3$), $u_z$ still
exhibits strong intermittency below $z\approx 0.1$ with
$\mbox{max}(\mathcal{K})\approx54$ whereas $\mathcal{K}$ for the
horizontal flows is significantly reduced in comparison to the
non-rotating case. This indicates that especially the vertical flows
in this regime are not qualitatively different from those in the
non-rotating regime, such that the downflows in the overshoot region
are rather abruptly decelerated and diverted horizontally. For the
most rapidly rotating case (Run~A9; $\Co=16.5$),
$\mathcal{K}\approx3\ldots4$ throughout the simulation domain for both
vertical and horizontal flows. This is explained by the almost
complete wiping out of the up-/downflow asymmetry also in the deep
parts of the CZ and in the overshoot region. The absence of a peak in
the kurtosis in the overshoot region in the most rapidly rotating
cases is likely due to the deeply penetrating vertical flows in those
cases due to the unrealistically small Richardson number. This is
discussed in more detail in \Sec{sec:OZDZ}.

\begin{figure}
  \includegraphics[width=\columnwidth]{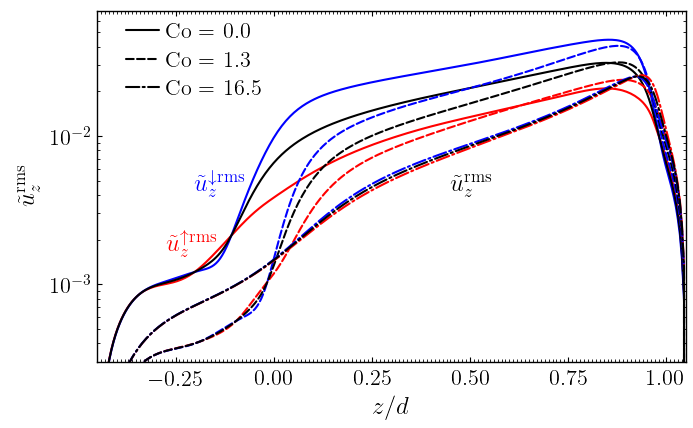}
\caption{Horizontally averaged vertical rms-velocity for the same runs
  as in \Fig{fig:plot_pdf}. The overall vertical velocity
  ($\tilde{u}_z^{\rm rms}$) is shown in black, and the corresponding
  quantities for up- ($\tilde{u}_z^{\uparrow{\rm rms}}$) and downflows
  ($\tilde{u}_z^{\downarrow{\rm rms}}$) are shown in red and blue,
  respectively. The tildes refers to normalization by $\sqrt{gd}$.}
\label{fig:puzrms}
\end{figure}

The average vertical rms-velocities from the same representative runs
as in \Fig{fig:plot_pdf} are shown in \Fig{fig:puzrms}. The average
rms-velocity of the downflows (upflows) is always larger (smaller)
than the average total vertical rms-velocity. However, the difference
between the up- and downflows and the total rms-velocity diminish
monotonically as a function of rotation such that for the most rapidly
rotating case the three are almost the same. This is another
manifestation of the symmetrization of up- and downflows.

\begin{figure}
  \includegraphics[width=\columnwidth]{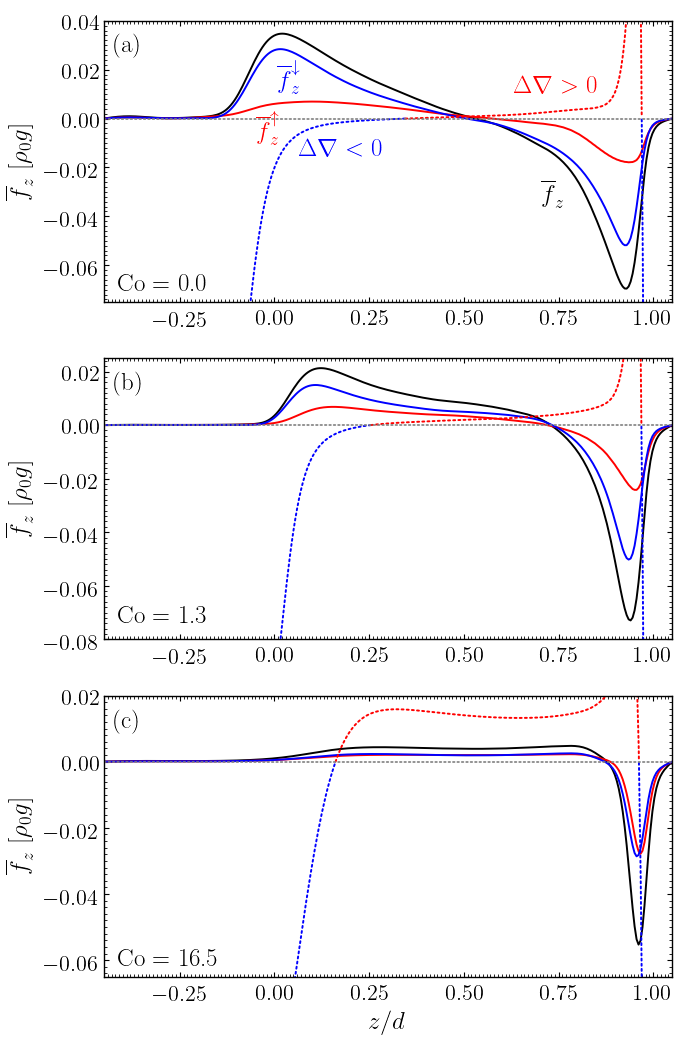}
\caption{Horizontally averaged total force (black), and separately for
  up- (red) and downflows (blue). The dotted red/blue line shows the
  superadiabatic temperature gradient. Data is shown for {\it (a)} a
  non-rotating run A0, {\it (b)} an intermediate rotation rate
  ($\Co=1.3$, Run~A6), and {\it (c)} for rapid rotation ($\Co=16.5$,
  Run~A9).}
\label{fig:plot_xyaver_forces}
\end{figure}

Another consequence of the symmetrization of the vertical flows is
that the forces on the up- and downflows also approach each other; see
\Figa{fig:plot_xyaver_forces}, where $\mean{f}_z = \mean{\rho
  Du_z/Dt}$. In accordance with earlier studies
\citep{2017ApJ...845L..23K,2019A&A...631A.122K}, in non-rotating
convection the downflows are accelerated near the surface and
decelerated roughly when the stratification turns Schwarzschild
stable, whereas the upflows are accelerated everywhere except near the
surface. This is interpreted such that the upflows are not driven by
buoyancy but by pressure forces due to the deeply penetrating downflow
plumes. This qualitative picture remains unchanged for slow rotation,
but starts to change when $\Co$ is of the order of unity although the
region near the surface where the downflows are accelerated is
shallower; see \Figa{fig:plot_xyaver_forces}(b). For rapid rotation
the forces on the up- and downflows are nearly identical. However, the
situation continues to qualitatively deviate from the mixing length
picture also in the rapidly rotating cases in that the downflows are
accelerated only near the surface and braked throughout their descent
through the superadiabatic CZ; see \Figa{fig:plot_xyaver_forces}(c).

\subsection{Overshooting and Deardorff layers}
\label{sec:OZDZ}

The depths of the overshooting and Deardorff layers are studied as
functions of rotation using the same definitions of overshooting and
Deardorff layers as in previous studies
\citep{2019A&A...631A.122K,2021A&A...655A..78K}. The bottom of the CZ
is situated at the depth $\zcz$ where $\mFconv$ changes from negative
to positive with increasing $z$. The top of the Deardorff zone (DZ) --
or the bottom of the buoyancy zone (BZ) -- $\zbz$, is where the
superadiabatic temperature gradient changes from negative to positive
with increasing $z$. Then the depth of the DZ is
\begin{eqnarray}
\ddz = \frac{1}{\Delta t} \int_{t_0}^{t_1} [\zbz(t)-\zcz(t)]dt,
\end{eqnarray}
where $\Delta t = t_1 - t_0$ is the length of the statistically steady
part of the time series. A reference value of the kinetic energy flux
($\mFkin^{\rm ref}$) is measured at $\zcz$. The base of the overshoot
layer is taken to be the location ($\zoskin$) where $|\mFkin|$ falls
below $0.01\mFkin^{\rm ref}$, and
\begin{eqnarray}
\doskin = \frac{1}{\Delta t} \int_{t_0}^{t_1} [\zcz(t)-\zoskin(t)]dt,\label{equ:dosk}
\end{eqnarray}
This criterion breaks down in the current models when rotation begins
to dominate the dynamics and where $\mFkin \rightarrow 0$. Therefore
the convected flux $\mFconv$ was also used to estimate the depth of
overshooting. The criterion involving $\mFconv$ takes the overshoot
layer to end at the location ($\zosconv$) where $|\mFconv|$ falls
below $0.02\Ftot$. The corresponding overshooting depth ($\dosconv$)
is computed analogously to \Eq{equ:dosk}. The layer below the OZ is
the radiative zone (RZ).

\begin{figure}
  \includegraphics[width=\columnwidth]{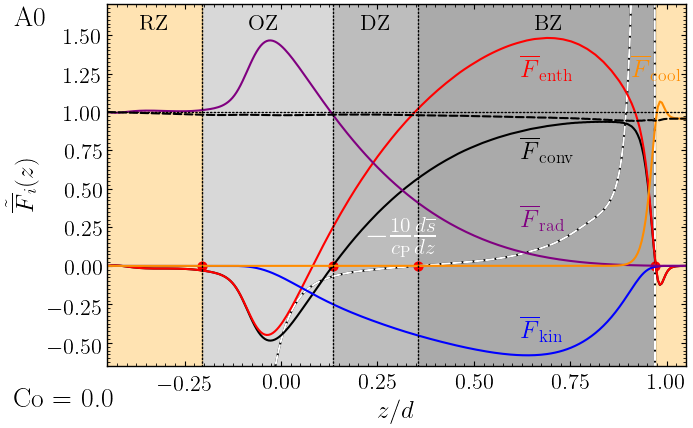}
  \includegraphics[width=\columnwidth]{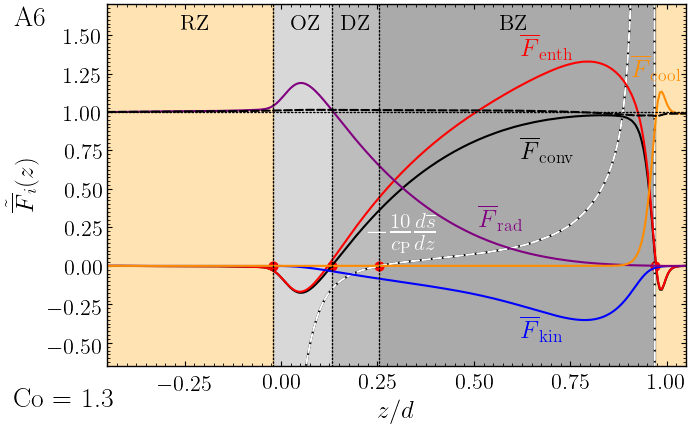}
  \includegraphics[width=\columnwidth]{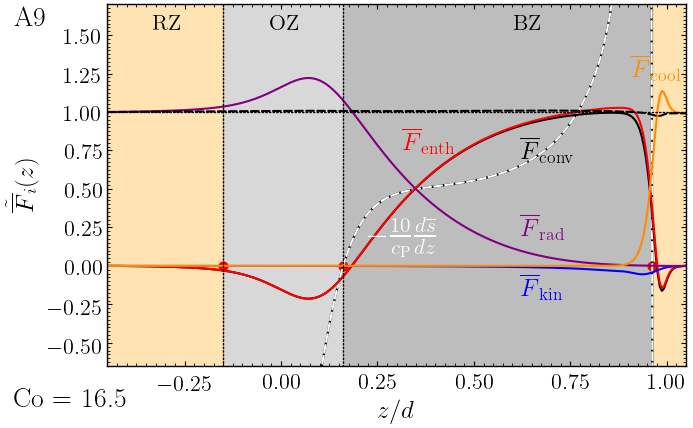}
\caption{Time-averaged mean energy fluxes as defined in
  \Equs{equ:mFrad}{equ:mFcool} (apart from the negligibly small
  viscous flux $\mFvisc$) as a functions of $z$ from Runs~A0, A6, and
  A9. The red circles indicate (from left to right) the bottoms of OZ,
  DZ, BZ, and the top of the BZ. The grey (orange) shaded areas
  indicate mixed (radiative) regions.}
\label{fig:plot_fluxes}
\end{figure}

\Fig{fig:plot_fluxes} shows the energy fluxes from representative runs
at different Coriolis numbers from Set~A. For slow and moderate
rotation up to $\Co\approx1$ the situation is qualitatively similar:
the positive (upward) enthalpy flux exceeds $\Fbot$ in the bulk of the
CZ, and it is compensated by a negative (downward) kinetic energy flux
$\mFkin$. As rotation increases the maxima of $\mFenth$ and $|\mFkin|$
decrease monotonically. Similarly, the extents of the overshoot and
Deardorff layers diminish with rotation. For the most rapidly rotation
case, Run~A9 with $\Co=16.5$, the kinetic energy flux is almost zero,
and $\mFconv\approx\mFenth$. This is yet another manifestation of the
decreasing asymmetry between the up- and downflows. Moreover, the
Deardorff layer vanishes in the rapidly rotating cases.

\begin{figure}
  \includegraphics[width=\columnwidth]{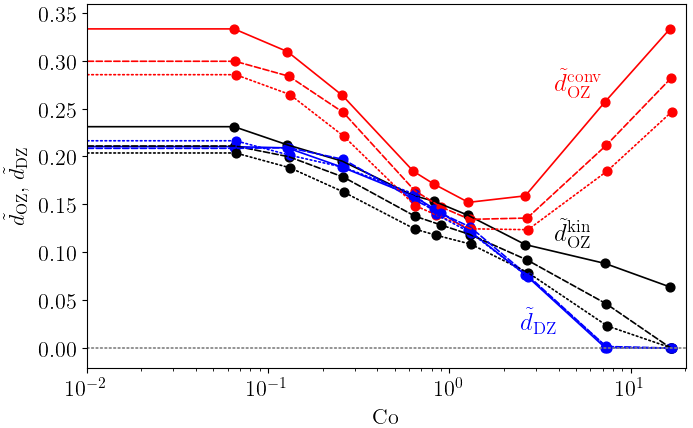}
\caption{Depth of the overshoot layer from kinetic energy ($\doskin$;
  black lines) and convective fluxes fluxes ($\dosconv$; red), and
  depth of the Deardorff layer ($\ddz$; blue) as functions of rotation
  measured by $\Co$. All quantities are normalized by the pressure
  scale height at the base of the CZ. The different lines correspond
  to the three different values of $\Fn$ or Sets~A (solid lines), B
  (dashed), and C (dotted).}
\label{fig:plot_oshoot_dear}
\end{figure}

The positions of the boundaries of the different layers and their
depths are summarized for all runs in \Table{tab:depths}, and
\Figa{fig:plot_oshoot_dear} shows a summary of the overshooting and
Deardorff layer depths as a function of rotation from Sets~A, B, and
C. The main difference between the sets of simulations of the applied
flux $\Fn$. The overshooting depth measured from the kinetic helicity
flux decreases with increasing rotation as in earlier studies
\citep[e.g.][]{2003A&A...401..433Z,KKT04}. However, the lowermost
panel of \Figa{fig:plot_fluxes} shows that the upper part of the
radiative layer is mixed far beyond the regions where $\mFkin$ is
non-negligible in the rapidly rotating cases. This is confirmed when
the convected flux is used to estimate the overshooting
depth. Furthermore, $\dosconv$ increases with rotation for $\Co
\gtrsim 1$. This is explained by the fact that the Mach number, and
therefore also the rotation rate $\Omega_0$, in the current
simulations are much larger than in real stars. This means that the
convective, rotation, and Brunt-V\"ais\"al\"a frequencies are closer
to each other in the simulations in comparison to, for example, the
overshoot region of the Sun. For example, in the most rapidly rotating
runs the Richardson number based on the rotation rate $\Ri_\Omega$ is
smaller than unity; see the 11th panel of \Table{tab:runs1}. This, in
addition to the smooth transition from convective to radiative region,
can lead gravity waves breaking in the radiative layer, thus
contributing to the burrowing of the flows into the RZ
\citep[e.g.][]{2013MNRAS.430.2363L}. As a comparison, $\Ri_\Omega$ in
the upper part of the solar radiative zone is expected of the order of
$10^{4}$. Another possibility is that shear due to the rotationally
constrained convective columns lowers the corresponding shear
Richardson number close to the limit where turbulence can occur also
in thermally stable stratification.

Lowering the luminosity in Sets~B and C shows that both measures of
$\dos$ decrease with $\Fn$ in qualitative accordance with earlier
results \citep[e.g.][]{2019A&A...631A.122K}. Even though $\Ri_\Omega$
is modestly increased in these runs (see the 11th column in
\Table{tab:runs1}), the most rapidly rotating cases even in the runs
with the lowest luminosities continue to show deep mixing which is
most likely due to the still unrealistically low $\Ri_\Omega$. It is
numerically very expensive to increase the Richardson number in fully
compressible simulations much further, at least without accelerated
thermal evolution methods
\citep[e.g.][]{2018PhRvF...3h3502A,2020PhRvF...5h3501A}. Comparing the
overshooting depths between Runs~[A,B,C]5 with solar $\CoF$ and the
non-rotating Runs~[A,B,C]0 shows a reduction between about a third to
a half; see the seventh and eight columns in \Table{tab:depths}. In
\cite{2019A&A...631A.122K} the overshooting depth extrapolated to the
solar value of $\Fn$ was found to be roughly $0.1\Hp$, and the current
results including rotation reduce this to $0.05\ldots
0.07\Hp$. However, the dependence of the overshooting depth on $\Fn$
is here steeper ($[\tdoskin,\tdosconv] \propto \Fn^{0.15}$) than in
the nonrotating cases where \cite{2019A&A...631A.122K} found
$\dos\propto\Fn^{0.08}$.

\begin{table}[t!]
\centering
\caption[]{Summary of the buoyancy, Deardorff, and overshoot zones.}
  \label{tab:depths}
      $$
          \begin{array}{p{0.065\linewidth}ccrcccc}
          \hline
          \hline
          \noalign{\smallskip}
Run  & \zbz/d  & \zdz/d  & \zoskin/d  &\zosconv/d  & \tddz & \tdoskin  & \tdosconv \\
\hline
 A0 & 0.355 & 0.134 & -0.096 & -0.204 & 0.221 & 0.230 & 0.338 \\
 A1 & 0.338 & 0.128 & -0.103 & -0.205 & 0.210 & 0.231 & 0.333 \\
 A2 & 0.333 & 0.124 & -0.088 & -0.185 & 0.209 & 0.212 & 0.309 \\
 A3 & 0.318 & 0.130 & -0.065 & -0.134 & 0.189 & 0.195 & 0.264 \\
 A4 & 0.290 & 0.131 & -0.028 & -0.054 & 0.159 & 0.159 & 0.185\\
 A5 & 0.278 & 0.132 & -0.021 & -0.039 & 0.146 & 0.153 & 0.171\\
 A6 & 0.255 & 0.131 & -0.007 & -0.021 & 0.123 & 0.138 & 0.152\\
 A7 & 0.211 & 0.134 &  0.026 & -0.025 & 0.077 & 0.108 & 0.159\\
 A8 & 0.154 & 0.154 &  0.065 & -0.103 & 0.001 & 0.088 & 0.257\\
 A9 & 0.161 & 0.183 &  0.120 & -0.150 & 0.000 & 0.064 & 0.333\\
\hline
 B0 & 0.338 & 0.124 & -0.094 & -0.185 & 0.214 & 0.218 & 0.309\\
 B1 & 0.329 & 0.121 & -0.090 & -0.179 & 0.208 & 0.211 & 0.299\\
 B2 & 0.326 & 0.117 & -0.082 & -0.166 & 0.209 & 0.200 & 0.284\\
 B3 & 0.321 & 0.124 & -0.054 & -0.122 & 0.197 & 0.179 & 0.246\\
 B4 & 0.280 & 0.125 & -0.012 & -0.039 & 0.154 & 0.138 & 0.164\\
 B5 & 0.264 & 0.124 & -0.005 & -0.024 & 0.140 & 0.128 & 0.147\\
 B6 & 0.252 & 0.125 &  0.007 & -0.009 & 0.127 & 0.119 & 0.134\\
 B7 & 0.204 & 0.128 &  0.036 & -0.008 & 0.076 & 0.092 & 0.136\\
 B8 & 0.138 & 0.136 &  0.090 & -0.075 & 0.002 & 0.046 & 0.212\\
 B9 & 0.133 & 0.156 &  0.156 & -0.126 & 0.000 & 0.000 & 0.281\\
\hline
 C0 & 0.323 & 0.116 & -0.086 & -0.166 & 0.206 & 0.203 & 0.283\\
 C1 & 0.336 & 0.119 & -0.084 & -0.166 & 0.216 & 0.204 & 0.285\\
 C2 & 0.316 & 0.115 & -0.074 & -0.150 & 0.201 & 0.188 & 0.265\\
 C3 & 0.304 & 0.116 & -0.047 & -0.105 & 0.189 & 0.163 & 0.221\\
 C4 & 0.278 & 0.118 & -0.007 & -0.031 & 0.160 & 0.124 & 0.149\\
 C5 & 0.259 & 0.119 &  0.001 & -0.020 & 0.140 & 0.118 & 0.139\\
 C6 & 0.240 & 0.119 &  0.011 & -0.005 & 0.120 & 0.108 & 0.124\\
 C7 & 0.196 & 0.121 &  0.043 & -0.002 & 0.074 & 0.079 & 0.124\\
 C8 & 0.129 & 0.129 &  0.106 & -0.056 & 0.000 & 0.023 & 0.185\\
 C9 & 0.118 & 0.140 &  0.140 & -0.106 & 0.000 & 0.000 & 0.247\\
\hline
A1m & 0.321 & 0.128 & -0.101 & -0.232 & 0.193 & 0.229 & 0.359\\
A3m & 0.309 & 0.130 & -0.056 & -0.143 & 0.178 & 0.187 & 0.274\\
A5m & 0.264 & 0.133 & -0.009 & -0.032 & 0.131 & 0.143 & 0.165\\
A6m & 0.250 & 0.131 & -0.002 & -0.017 & 0.119 & 0.133 & 0.148\\
A7m & 0.218 & 0.133 &  0.018 & -0.015 & 0.085 & 0.115 & 0.148\\
A8m & 0.165 & 0.147 &  0.064 & -0.093 & 0.018 & 0.083 & 0.240\\
A9m & 0.163 & 0.177 &  0.052 & -0.163 & 0.000 & 0.126 & 0.340\\
\hline
A5h & 0.254 & 0.134 & -0.010 & -0.031 & 0.120 & 0.144 & 0.165\\
A9h & 0.170 & 0.170 &  0.060 & -0.183 & 0.000 & 0.110 & 0.353\\
\hline
          \end{array}
          $$ \tablefoot{The tildes for refer to normalization by the
            pressure scale height at the base of the convection zone.}
\end{table}

On the other hand, the thickness of the Deardorff layer $\ddz$
decreases monotonously as a function of $\Co$. In the most rapidly
rotating cases the Deardorff layer vanishes altogether and even
reverses such that at the base of the CZ the stratification is
unstably stratified but the convective flux is inward; see the
lowermost panel of \Fig{fig:plot_fluxes}. This is not significantly
changed in more supercritical Runs~A9m and A9h. In the entropy rain
picture \citep[e.g.][]{Br16} cool material from the surface is brought
down deep into otherwise stably stratified layers. This is mediated by
relatively few fast downflows with filling factor $f(z) < 1/2$, that
also produce a strong net downward kinetic energy flux as seen in the
top panel of \Fig{fig:plot_fluxes}; see also \Figa{fig:plot_ff}, and
Table~1 and Sect.~3.3 in \cite{Br16}. If, on the other hand, the up-
and downflows are symmetrized such that $f(z)=1/2$ and their
velocities are nearly the same, $\mFkin$ vanishes and non-local
transport due to downflows is no longer significant. Therefore the
kinetic energy flux is a proxy of the non-local transport due to
downflows and its absence signifies the absence of a Deardorff layer.
The depth of the Deardorff layer is independent of the energy flux
$\Fn$. This further illustrates that the DZ is caused by surface
effects which are kept independent of $\Fn$ in the current
simulations. A reduction of $\ddz$ of about a third between the
non-rotating runs [A,B,C]0 and the runs with the solar value of $\CoF$
(Runs~[A,B,C]5) was found; see the sixth column of \Table{tab:depths}.

\section{Conclusions} \label{sect:conclusions}

Simulations of compressible convection were used to study the
convective scale and scalings of quantitites such as the Coriolis
number and convective velocity as functions of rotation. The results
were compared to those expected from scalings obtained for
incompressible convection with slow and fast rotation
\citep{2020PhRvR...2d3115A}. The actual length scale is almost
unaffected by rotation for $\Co\lesssim 1$ and decreases proportional
to $\Co^{1/2}$ for rapid rotation. Correspondingly, the dynamical
Coriolis number $\Col$ is proportional to $\Co$ for slow, and
$\propto\Co^{1/2}$ for rapid rotation. Furthermore, $\Col$ is
proportional to $(\RaFS)^{-1/3}$ for slow and $\propto(\RaFS)^{-1/5}$
for rapid rotation, where $\RaFS$ is the diffusion-free flux-based
modified Rayleigh number. Finally, the convective velocity is
compatible with proportionality to $(\Ftot/\rho)^{1/3}$ for slow and
$\propto(\Ftot/\rho)^{1/3}\Co^{-1/6}$ for rapid rotation. All of these
scalings are consistent with those derived by
\cite{2020PhRvR...2d3115A} and \cite{2021PNAS..11822518V}. Therefore
the simulations seem to follow the CIA scaling at sufficiently rapid
rotation.

In an earlier work \citep{2023A&A...669A..98K} several measures were
used to characterise the rotational influence on convection. A
commonly used definition where the changing length scale of convection
is taken into account is $\Cow = 2\Omega/\orms$. It is shown that this
quantity cannot be used to characterise the effects of rotation on the
mean scale because $\orms$ is expected to increase with the Reynolds
number as $\Rey^{1/2}$. Therefore the only reliable way to account for
the changing convective length scale as a function of rotation is to
compute the mean wavenumber. This was not correctly identified in
\cite{2023A&A...669A..98K}, and it is now clear that $\Cow$ will
diverge as $\Rey$ increases. On the other hand,
\cite{2023A&A...669A..98K} introduced a stellar Coriolis number
$\Costar$ which depends on luminosity and rotation rate which are
observable and a reference density which is available from stellar
structure models, but not on any dynamical lenght or velocity
scale. Here this quantity is renamed as $\CoF$ and it is furthermore
shown that with a suitable choice of length scale,
$\CoF=(\RaFS)^{-1/3}$. Matching $\CoF$ (or equivalently $\RaFS$) with
the target star gives a more concrete meaning to the often-used phrase
that it is possible to match the Coriolis number of, for example, the
Sun with 3D simulations while most other dimensionless parameters are
out of reach \citep[cf.][]{2023SSRv..219...58K}.

The current simulations suggest that convection even in the deep parts
of the CZ in the Sun is not strongly rotationally constrained and that
the CIA balance is therefore inapplicable there. The latter has been
argued to be the case by \cite{FH16b} and \cite{2021PNAS..11822518V}
to argue that the largest convectively driven scale in the Sun is the
supergranular scale. The current results seem to refute this
conjecture and that the actual scales may be larger.

Finally, the effects of rotation on convective overshooting and
subadiabatic Deardorff zones were studied. The effects of rotation are
relatively mild such that for the case with the solar value of $\CoF$,
the overshooting depth and the extent of the Deardorff layer are
reduced by between 30 and 50 per cent in comparison to the
non-rotating case. Therefore the current results suggest an
overshooting depth of about five per cent of the pressure scale height
at the base of the solar CZ. Taking the current results at face value,
a similar depth is estimated for the Deardorf zone. However, the
latter is still subject to the caveat that the current simulations do
not capture the near-surface layer very accurately and that the
driving of entropy rain can be significantly stronger in
reality. Another aspect which needs to be revisited in the future is
the effect of magnetic fields.

\begin{acknowledgements}
  I thank Axel Brandenburg for his comments on an earlier version of
  the manuscript. The simulations were performed using the resources
  granted by the Gauss Center for Supercomputing for the Large-Scale
  computing project ``Cracking the Convective Conundrum'' in the
  Leibniz Supercomputing Centre's SuperMUC-NG supercomputer in
  Garching, Germany. This work was supported in part by the Deutsche
  Forschungsgemeinschaft Heisenberg programme (grant No.\ KA
  4825/4-1).
\end{acknowledgements}

\bibliographystyle{aa}
\bibliography{paper}

\appendix

\section{Convective scalings as function of rotation}
\label{app:convsca}

The scalings of relevant dynamical quantities in convection are
shortly summarized following the studies of
\cite{2014ApJ...791...13B}, \cite{2020PhRvR...2d3115A}, and
\cite{2021PNAS..11822518V}. In the rapidly rotating regime the
Coriolis-inertial-Archimedean (CIA) balance is assumed to hold.

\subsection{No or slow rotation ($\Co\ll1$)}

For slow rotation the convective length scale $\lconv$ is of the order
of the pressure scale height $\Hp=-(\pd\ln p/\pd z)^{-1}$, and the
vertical and horizontal extents of convection cells are of the same
order of magnitude. When rotation is slow, the dominant balance in the
Navier--Stokes equation is between the advection and buoyancy terms:
\begin{eqnarray}
\uuu\bm\cdot\bm\nabla\uuu \sim \frac{T'}{T}\gggg\ \longrightarrow \frac{u^2}{H}
 \sim \frac{T'}{T} g,
\end{eqnarray}
where $H$ is the convective scale $T'$ is the temperature
fluctuation. Assuming that convection transports most of the energy
gives
\begin{eqnarray}
\Ftot \sim \cP \rho u T',
\end{eqnarray}
and therefore
\begin{eqnarray}
u^3 \sim \frac{gH}{\cP T}\frac{\Ftot}{\rho},\ \mbox{or}\ \ u \sim
\left(\frac{g H}{\cP
  T}\right)^{1/3}\left(\frac{\Ftot}{\rho}\right)^{1/3}.
\end{eqnarray}
Choosing $H = \cP T/g$ gives:
\begin{eqnarray}
u \sim \left(\frac{\Ftot}{\rho}\right)^{1/3} \equiv \ustar,\label{equ:unorot}
\end{eqnarray}
where $\ustar$ is a hypothetical velocity that is a measure of the
available energy flux. Therefore, for slow rotation,
\begin{eqnarray}
\Co = \Col = \frac{2\Omega H}{u} = 2\Omega H \left(\frac{\rho}{\Ftot} \right)^{1/3} = (\RaFS)^{-1/3}.
\end{eqnarray}
Temperature fluctuation can be computed from the convective flux
\begin{eqnarray}
  \Fconv = \cP \rho u T',\ \ \longrightarrow\ \ \cP T' = \frac{F}{\rho u}.\label{equ:TprimeFconv}
\end{eqnarray}
Using $u$ from \Eq{equ:unorot} yields:
\begin{eqnarray}
\cP T' = \left(\frac{F}{\rho}\right)^{2/3}.
\end{eqnarray}

\subsection{Rapid rotation ($\Co\gg1$)}

The CIA balance means that
\begin{eqnarray}
2\Omega_0\pd_\parallel \uuu \sim \uuu\bm\cdot\bm\nabla \ooo \sim \bm\nabla \times \left( \frac{T'}{T} \gggg\right),
\end{eqnarray}
which results from the curl of the Navier--Stokes equation
\citep[e.g.][]{2020PhRvR...2d3115A}, and where $\pd_\parallel$ is a
derivative along the rotation vector. Considering first the
CI part of CIA balance gives
\begin{eqnarray}
  \uuu\bm\cdot\bm\nabla\ooo \sim
  2\Omega\pd_z\uuu\ \ \longrightarrow\ \ k_\perp^2 u^2 \sim 2\Omega u
  k_\parallel,
\end{eqnarray}
where $k_\perp$ and $k_\parallel$ are the wavenumbers perpendicular
and parallel to the rotation vector. Identifying $k_\perp$ as the
dominant horizontal scale of convection ($\kmax\sim \ell^{-1}$) and
$k_\parallel$ as $k_1\sim H^{-1}$, leads to
\begin{eqnarray}
\left(\frac{H}{\ell}\right)^2 \propto \left(\frac{\kmax}{k_1}\right)^2 \propto \frac{2\Omega}{k_1 u} = \Co,\ \ \mbox{or}\ \ \frac{\ell}{H} = \Co^{-1/2}. \label{equ:kmaxk1}
\end{eqnarray}
\noindent
Furthermore,
\begin{eqnarray}
\Co = \frac{2\Omega H}{u} = \frac{2\Omega \ell}{u} \frac{H}{\ell} = \Col \Co^{1/2},
\end{eqnarray}
or
\begin{eqnarray}
\Col = \Co^{1/2}.\label{equ:colco}
\end{eqnarray}
\noindent
The convective length scale in terms of $u$ and global quantities is
\begin{eqnarray}
\ell = \left(\frac{Hu}{2\Omega}\right)^{1/2}.\label{equ:ellCol}
\end{eqnarray}
To derive the convective velocity, CA part of the CIA balance is used:
\begin{eqnarray}
  2\Omega_0\pd_\parallel \uuu \sim \bm\nabla \times \left(\frac{T'}{T}\gggg \right),\ \ \longrightarrow\ \ \frac{2\Omega u}{H} \sim \frac{g\Ftot}{\cP \rho T u \ell}. \label{equ:U2CIA}
\end{eqnarray}
Substitute $\ell$ from \Eq{equ:ellCol} and rearrange to get:
\begin{eqnarray}
u = \left(\frac{g\Ftot}{\cP \rho T}\right)^{2/5} \left(\frac{H}{2\Omega} \right)^{1/5} = \left(\frac{\Ftot}{\rho}\right)^{2/5} (2\Omega H)^{-1/5},\label{equ:UCIA}
\end{eqnarray}
where $H=\cP T/g$ was additionally used.
This is equivalent to:
\begin{eqnarray}
u = \left(\frac{\Ftot}{\rho}\right)^{1/3} \Co^{-1/6}.\label{equ:UfCo}
\end{eqnarray}
The length scale $\ell$ is obtained from \Eq{equ:U2CIA} with
substitution of $u$ from \Eq{equ:UCIA}:
\begin{eqnarray}
  \frac{\ell}{H} = \left(\frac{\Ftot}{\rho}\right)^{1/5} (2\Omega H)^{-3/5},
\end{eqnarray}
where $H=\cP T/g$ was again used. Now,
\begin{eqnarray}
\Col = \frac{2\Omega\ell}{u} = (2\Omega H)^{3/5} \left(\frac{\rho}{\Ftot}\right)^{1/5}.
\end{eqnarray}
Bearing \Eq{equ:RaFS_simple} in mind gives:
\begin{eqnarray}
\Col = \left(\frac{8\Omega^3 H^3 \rho}{F} \right)^{1/5} = (\RaFS)^{-1/5}.
\end{eqnarray}
\noindent
Finally, the temperature fluctuation using \Eq{equ:UfCo} is:
\begin{eqnarray}
\cP T' = \frac{\Ftot}{\rho u} = \left(\frac{\Ftot}{\rho}\right)^{2/3} \Co^{1/6}.
\end{eqnarray}

\end{document}

%% file: macros.tex
\usepackage{xcolor}

\newcommand{\uuu}{{\bm u}}

\newcommand{\xxx}{{\bm x}}

\newcommand{\ooo}{{\bm \omega}}

\newcommand{\FFF}{{\bm F}}

\newcommand{\gggg}{{\bm g}}

\newcommand{\muuz}{\overline{u}_z}

\newcommand{\SSt}{\bm{\mathsf{S}}}
\newcommand{\SStij}{\mathsf{S}_{ij}}

\newcommand{\Eq}[1]{Eq.~(\ref{#1})}
\newcommand{\Equ}[1]{Equation~(\ref{#1})}

\newcommand{\Equs}[2]{Equations~(\ref{#1}) to (\ref{#2})}

\newcommand{\EQ}{\begin{equation}}
\newcommand{\EN}{\end{equation}}
\newcommand{\EQA}{\begin{eqnarray}}
\newcommand{\ENA}{\end{eqnarray}}

\newcommand{\pd}{\partial}

\newcommand{\mean}[1]{\overline{#1}}

\newcommand{\cP}{c_{\rm P}}
\newcommand{\cV}{c_{\rm V}}

\newcommand{\urms}{u_{\rm rms}}
\newcommand{\uconv}{u_{\rm conv}}
\newcommand{\ustar}{u_\star}
\newcommand{\uflux}{u_{\rm flux}}

\newcommand{\orms}{\omega_{\rm rms}}

\newcommand{\Ma}{{\rm Ma}}

\newcommand{\kmax}{k_{\rm max}}

\newcommand{\kmean}{k_{\rm mean}}
\newcommand{\mkmean}{\mean{k}_{\rm mean}}

\newcommand{\lconv}{\ell_{\rm conv}}
\newcommand{\chiSGS}{\chi_{\rm SGS}}

\newcommand{\chieff}{\chi_{\rm eff}}
\newcommand{\mchieff}{\overline{\chi}_{\rm eff}}

\newcommand{\Co}{{\rm Co}}

\newcommand{\Costar}{{\rm Co}_\star}
\newcommand{\Cow}{{\rm Co}_\omega}
\newcommand{\Col}{{\rm Co}_\ell}
\newcommand{\CoF}{{\rm Co}_{\rm F}}

\newcommand{\Ek}{{\rm Ek}}

\newcommand{\Hp}{H_{\rm p}}

\newcommand{\Pe}{{\rm Pe}}
\newcommand{\Pel}{{\rm Pe}_\ell}
\newcommand{\PeSGS}{{\rm Pe}_{\rm SGS}}
\newcommand{\Pra}{{\rm Pr}}

\newcommand{\PraSGS}{{\rm Pr}_{\rm SGS}}

\newcommand{\Ra}{{\rm Ra}}
\newcommand{\Rat}{{\rm Ra}_{\rm t}}
\newcommand{\RaF}{{\rm Ra}_{\rm F}}
\newcommand{\RaFS}{{\rm Ra}_{\rm F}^\star}

\newcommand{\Rey}{{\rm Re}}
\newcommand{\Rel}{{\rm Re}_\ell}

\newcommand{\Ro}{{\rm Ro}}

\newcommand{\Ta}{{\rm Ta}}
\newcommand{\Ri}{{\rm Ri}}
\newcommand{\taucool}{\tau_{\rm cool}}

{}

\newcommand{\calR}{{\cal R}}

\newcommand{\Omsun}{\Omega_\odot}

\newcommand{\rhostar}{\rho_\star}

\newcommand{\Rsun}{R_\odot}

\newcommand{\Fconv}{F_{\rm conv}}

\newcommand{\Fbot}{F_{\rm bot}}
\newcommand{\Ftot}{F_{\rm tot}}

\newcommand{\Fn}{\mathscr{F}_{\rm n}}
\newcommand{\mFenth}{\mean{F}_{\rm enth}}
\newcommand{\mFconv}{\mean{F}_{\rm conv}}
\newcommand{\mFrad}{\mean{F}_{\rm rad}}
\newcommand{\mFkin}{\mean{F}_{\rm kin}}
\newcommand{\mFvisc}{\mean{F}_{\rm visc}}

\newcommand{\mFcool}{\mean{F}_{\rm cool}}

\newcommand{\Lsun}{L_\odot}
\newcommand{\nabad}{\nabla_{\rm ad}}

\newcommand{\tbot}{{\rm bot}}

\newcommand{\zbot}{z_{\rm bot}}
\newcommand{\zbz}{z_{\rm BZ}}
\newcommand{\zcz}{z_{\rm CZ}}
\newcommand{\zdz}{z_{\rm DZ}}

\newcommand{\zosconv}{z_{\rm OS}^{\rm conv}}

\newcommand{\zoskin}{z_{\rm OS}^{\rm kin}}

\newcommand{\ddz}{d_{\rm DZ}}
\newcommand{\tddz}{\tilde{d}_{\rm DZ}}
\newcommand{\dos}{d_{\rm os}}

\newcommand{\dosconv}{d_{\rm os}^{\rm conv}}
\newcommand{\doskin}{d_{\rm os}^{\rm kin}}
\newcommand{\tdosconv}{\tilde{d}_{\rm os}^{\rm conv}}
\newcommand{\tdoskin}{\tilde{d}_{\rm os}^{\rm kin}}
\newcommand{\nrad}{n_{\rm rad}}
\newcommand{\nad}{n_{\rm ad}}
\def\onethird{{\textstyle{1\over3}}}

\def\onehalf{{\textstyle{1\over2}}}

\newcommand{\mepsK}{\overline{\epsilon}_{\rm K}}

\newcommand{\Figa}[1]{Fig.~\ref{#1}}

\newcommand{\Fig}[1]{Figure~\ref{#1}} 
\newcommand{\Figu}[1]{Figure~\ref{#1}}

\newcommand{\Sec}[1]{Section~\ref{#1}} 
\newcommand{\Secs}[2]{Sections~\ref{#1} and \ref{#2}}
\newcommand{\Table}[1]{Table~\ref{#1}}

\newcommand{\Appendix}[1]{Appendix~\ref{#1}}

\definecolor{ForestGreen}{RGB}{34,139,34}
\definecolor{AGray}{rgb}{.4,.4,.4}
\definecolor{LightYellow}{rgb}{1.,1.,.8}
\definecolor{LightCyan}{rgb}{0.88,1,1}

%% file: paper.bbl
\begin{thebibliography}{75}
\expandafter\ifx\csname natexlab\endcsname\relax\def\natexlab#1{#1}\fi

\bibitem[{{Anders} {et~al.}(2018){Anders}, {Brown}, \&
  {Oishi}}]{2018PhRvF...3h3502A}
{Anders}, E.~H., {Brown}, B.~P., \& {Oishi}, J.~S. 2018, Physical Review
  Fluids, 3, 083502

\bibitem[{{Anders} {et~al.}(2022){Anders}, {Jermyn}, {Lecoanet}, \&
  {Brown}}]{2022ApJ...926..169A}
{Anders}, E.~H., {Jermyn}, A.~S., {Lecoanet}, D., \& {Brown}, B.~P. 2022, \apj,
  926, 169

\bibitem[{{Anders} \& {Pedersen}(2023)}]{2023Galax..11...56A}
{Anders}, E.~H. \& {Pedersen}, M.~G. 2023, Galaxies, 11, 56

\bibitem[{{Anders} {et~al.}(2020){Anders}, {Vasil}, {Brown}, \&
  {Korre}}]{2020PhRvF...5h3501A}
{Anders}, E.~H., {Vasil}, G.~M., {Brown}, B.~P., \& {Korre}, L. 2020, Physical
  Review Fluids, 5, 083501

\bibitem[{{Aurnou} {et~al.}(2020){Aurnou}, {Horn}, \&
  {Julien}}]{2020PhRvR...2d3115A}
{Aurnou}, J.~M., {Horn}, S., \& {Julien}, K. 2020, Physical Review Research, 2,
  043115

\bibitem[{{Barekat} \& {Brandenburg}(2014)}]{BB14}
{Barekat}, A. \& {Brandenburg}, A. 2014, \aap, 571, A68

\bibitem[{{Barker} {et~al.}(2014){Barker}, {Dempsey}, \&
  {Lithwick}}]{2014ApJ...791...13B}
{Barker}, A.~J., {Dempsey}, A.~M., \& {Lithwick}, Y. 2014, \apj, 791, 13

\bibitem[{{Bekki} {et~al.}(2017){Bekki}, {Hotta}, \&
  {Yokoyama}}]{2017ApJ...851...74B}
{Bekki}, Y., {Hotta}, H., \& {Yokoyama}, T. 2017, \apj, 851, 74

\bibitem[{{B{\"o}hm-Vitense}(1958)}]{BV58}
{B{\"o}hm-Vitense}, E. 1958, \zap, 46, 108

\bibitem[{{Brandenburg}(2016)}]{Br16}
{Brandenburg}, A. 2016, \apj, 832, 6

\bibitem[{{Brandenburg} {et~al.}(2005){Brandenburg}, {Chan}, {Nordlund}, \&
  {Stein}}]{BCNS05}
{Brandenburg}, A., {Chan}, K.~L., {Nordlund}, {\AA}., \& {Stein}, R.~F. 2005,
  AN, 326, 681

\bibitem[{{Brandenburg} {et~al.}(1996){Brandenburg}, {Jennings}, {Nordlund},
  {Rieutord}, {Stein}, \& {Tuominen}}]{BJNRST96}
{Brandenburg}, A., {Jennings}, R.~L., {Nordlund}, {\AA}., {et~al.} 1996, J.
  Fluid Mech., 306, 325

\bibitem[{{Brandenburg} {et~al.}(2000){Brandenburg}, {Nordlund}, \&
  {Stein}}]{2000gac..conf...85B}
{Brandenburg}, A., {Nordlund}, A., \& {Stein}, R.~F. 2000, in Geophysical and
  Astrophysical Convection, Contributions from a workshop sponsored by the
  Geophysical Turbulence Program at the National Center for Atmospheric
  Research, October, 1995. Edited by Peter A. Fox and Robert M. Kerr. Published
  by Gordon and Breach Science Publishers, The Netherlands, 2000, p. 85-105,
  ed. P.~A. {Fox} \& R.~M. {Kerr}, 85--105

\bibitem[{{Brandenburg} \& {Petrosyan}(2012)}]{2012AN....333..195B}
{Brandenburg}, A. \& {Petrosyan}, A. 2012, Astronomische Nachrichten, 333, 195

\bibitem[{{Brummell} {et~al.}(2002){Brummell}, {Clune}, \& {Toomre}}]{BCT02}
{Brummell}, N.~H., {Clune}, T.~L., \& {Toomre}, J. 2002, \apj, 570, 825

\bibitem[{{Brun} {et~al.}(2017){Brun}, {Strugarek}, {Varela}, {Matt},
  {Augustson}, {Emeriau}, {DoCao}, {Brown}, \& {Toomre}}]{2017ApJ...836..192B}
{Brun}, A.~S., {Strugarek}, A., {Varela}, J., {et~al.} 2017, \apj, 836, 192

\bibitem[{{Candelaresi} \& {Brandenburg}(2013)}]{CB13}
{Candelaresi}, S. \& {Brandenburg}, A. 2013, \pre, 87, 043104

\bibitem[{{Cattaneo} {et~al.}(1991){Cattaneo}, {Brummell}, {Toomre},
  {Malagoli}, \& {Hurlburt}}]{CBTMH91}
{Cattaneo}, F., {Brummell}, N.~H., {Toomre}, J., {Malagoli}, A., \& {Hurlburt},
  N.~E. 1991, \apj, 370, 282

\bibitem[{{Chan}(2003)}]{Chan03}
{Chan}, K.~L. 2003, in Astron. Soc. Pac. Conf. Ser., Vol. 293, 3D Stellar
  Evolution, ed. {S.~Turcotte, S.~C.~Keller, \& R.~M.~Cavallo}, 168

\bibitem[{{Chan}(2007)}]{Chan07}
{Chan}, K.~L. 2007, Astron. Nachr., 328, 1059

\bibitem[{{Chan} \& {Mayr}(2013)}]{2013E&PSL.371..212C}
{Chan}, K.~L. \& {Mayr}, H.~G. 2013, Earth and Planetary Science Letters, 371,
  212

\bibitem[{{Chandrasekhar}(1961)}]{Ch61}
{Chandrasekhar}, S. 1961, {Hydrodynamic and hydromagnetic stability}

\bibitem[{{Christensen}(2002)}]{2002JFM...470..115C}
{Christensen}, U.~R. 2002, Journal of Fluid Mechanics, 470, 115

\bibitem[{{Christensen} \& {Aubert}(2006)}]{CA06}
{Christensen}, U.~R. \& {Aubert}, J. 2006, Geophys. J. Int., 166, 97

\bibitem[{{Currie} {et~al.}(2020){Currie}, {Barker}, {Lithwick}, \&
  {Browning}}]{2020MNRAS.493.5233C}
{Currie}, L.~K., {Barker}, A.~J., {Lithwick}, Y., \& {Browning}, M.~K. 2020,
  \mnras, 493, 5233

\bibitem[{{Deardorff}(1961)}]{1961JAtS...18..540D}
{Deardorff}, J.~W. 1961, J. Atmosph. Sci., 18, 540

\bibitem[{{Deardorff}(1966)}]{De66}
{Deardorff}, J.~W. 1966, J. Atmosph. Sci., 23, 503

\bibitem[{{Dobler} {et~al.}(2006){Dobler}, {Stix}, \& {Brandenburg}}]{DSB06}
{Dobler}, W., {Stix}, M., \& {Brandenburg}, A. 2006, \apj, 638, 336

\bibitem[{{Edwards}(1990)}]{1990MNRAS.242..224E}
{Edwards}, J.~M. 1990, \mnras, 242, 224

\bibitem[{{Featherstone} \& {Hindman}(2016)}]{FH16b}
{Featherstone}, N.~A. \& {Hindman}, B.~W. 2016, \apjl, 830, L15

\bibitem[{{Gastine} {et~al.}(2014){Gastine}, {Yadav}, {Morin}, {Reiners}, \&
  {Wicht}}]{GYMRW14}
{Gastine}, T., {Yadav}, R.~K., {Morin}, J., {Reiners}, A., \& {Wicht}, J. 2014,
  \mnras, 438, L76

\bibitem[{{Greer} {et~al.}(2015){Greer}, {Hindman}, {Featherstone}, \&
  {Toomre}}]{GHFT15}
{Greer}, B.~J., {Hindman}, B.~W., {Featherstone}, N.~A., \& {Toomre}, J. 2015,
  \apjl, 803, L17

\bibitem[{{Guervilly} {et~al.}(2014){Guervilly}, {Hughes}, \& {Jones}}]{GHJ14}
{Guervilly}, C., {Hughes}, D.~W., \& {Jones}, C.~A. 2014, J. Fluid Mech., 758,
  407

\bibitem[{{Hanasoge} {et~al.}(2016){Hanasoge}, {Gizon}, \&
  {Sreenivasan}}]{2016AnRFM..48..191H}
{Hanasoge}, S., {Gizon}, L., \& {Sreenivasan}, K.~R. 2016, Annual Review of
  Fluid Mechanics, 48, 191

\bibitem[{{Hanasoge} {et~al.}(2012){Hanasoge}, {Duvall}, \&
  {Sreenivasan}}]{HDS12}
{Hanasoge}, S.~M., {Duvall}, T.~L., \& {Sreenivasan}, K.~R. 2012, Proc. Natl.
  Acad. Sci., 109, 11928

\bibitem[{{Hotta}(2017)}]{2017ApJ...843...52H}
{Hotta}, H. 2017, \apj, 843, 52

\bibitem[{{Hotta} {et~al.}(2015){Hotta}, {Rempel}, \& {Yokoyama}}]{HRY15b}
{Hotta}, H., {Rempel}, M., \& {Yokoyama}, T. 2015, \apj, 803, 42

\bibitem[{{Ingersoll} \& {Pollard}(1982)}]{1982Icar...52...62I}
{Ingersoll}, A.~P. \& {Pollard}, D. 1982, \icarus, 52, 62

\bibitem[{{K{\"a}pyl{\"a}}(2019)}]{2019A&A...631A.122K}
{K{\"a}pyl{\"a}}, P.~J. 2019, \aap, 631, A122

\bibitem[{{K{\"a}pyl{\"a}}(2021)}]{2021A&A...655A..78K}
{K{\"a}pyl{\"a}}, P.~J. 2021, \aap, 655, A78

\bibitem[{{K{\"a}pyl{\"a}}(2023)}]{2023A&A...669A..98K}
{K{\"a}pyl{\"a}}, P.~J. 2023, \aap, 669, A98

\bibitem[{{K{\"a}pyl{\"a}} {et~al.}(2023){K{\"a}pyl{\"a}}, {Browning}, {Brun},
  {Guerrero}, \& {Warnecke}}]{2023SSRv..219...58K}
{K{\"a}pyl{\"a}}, P.~J., {Browning}, M.~K., {Brun}, A.~S., {Guerrero}, G., \&
  {Warnecke}, J. 2023, \ssr, 219, 58

\bibitem[{{K{\"a}pyl{\"a}} {et~al.}(2020){K{\"a}pyl{\"a}}, {Gent}, {Olspert},
  {K{\"a}pyl{\"a}}, \& {Brandenburg}}]{2020GApFD.114....8K}
{K{\"a}pyl{\"a}}, P.~J., {Gent}, F.~A., {Olspert}, N., {K{\"a}pyl{\"a}}, M.~J.,
  \& {Brandenburg}, A. 2020, Geophysical and Astrophysical Fluid Dynamics, 114,
  8

\bibitem[{{K{\"a}pyl{\"a}} {et~al.}(2004){K{\"a}pyl{\"a}}, {Korpi}, \&
  {Tuominen}}]{KKT04}
{K{\"a}pyl{\"a}}, P.~J., {Korpi}, M.~J., \& {Tuominen}, I. 2004, \aap, 422, 793

\bibitem[{{K{\"a}pyl{\"a}} {et~al.}(2011){K{\"a}pyl{\"a}}, {Mantere}, \&
  {Hackman}}]{2011ApJ...742...34K}
{K{\"a}pyl{\"a}}, P.~J., {Mantere}, M.~J., \& {Hackman}, T. 2011, \apj, 742, 34

\bibitem[{{K{\"a}pyl{\"a}} {et~al.}(2017){K{\"a}pyl{\"a}}, {Rheinhardt},
  {Brandenburg}, {Arlt}, {K{\"a}pyl{\"a}}, {Lagg}, {Olspert}, \&
  {Warnecke}}]{2017ApJ...845L..23K}
{K{\"a}pyl{\"a}}, P.~J., {Rheinhardt}, M., {Brandenburg}, A., {et~al.} 2017,
  \apjl, 845, L23

\bibitem[{{K{\"a}pyl{\"a}} {et~al.}(2019){K{\"a}pyl{\"a}}, {Viviani},
  {K{\"a}pyl{\"a}}, {Brandenburg}, \& {Spada}}]{2019GApFD.113..149K}
{K{\"a}pyl{\"a}}, P.~J., {Viviani}, M., {K{\"a}pyl{\"a}}, M.~J., {Brandenburg},
  A., \& {Spada}, F. 2019, Geophysical and Astrophysical Fluid Dynamics, 113,
  149

\bibitem[{{Karak} {et~al.}(2018){Karak}, {Miesch}, \&
  {Bekki}}]{2018PhFl...30d6602K}
{Karak}, B.~B., {Miesch}, M., \& {Bekki}, Y. 2018, Physics of Fluids, 30,
  046602

\bibitem[{{King} \& {Buffett}(2013)}]{2013E&PSL.371..156K}
{King}, E.~M. \& {Buffett}, B.~A. 2013, Earth and Planetary Science Letters,
  371, 156

\bibitem[{{Kupka} \& {Muthsam}(2017)}]{2017LRCA....3....1K}
{Kupka}, F. \& {Muthsam}, H.~J. 2017, Liv. Rev. Comp. Astrophys., 3, 1

\bibitem[{{Lecoanet} \& {Quataert}(2013)}]{2013MNRAS.430.2363L}
{Lecoanet}, D. \& {Quataert}, E. 2013, \mnras, 430, 2363

\bibitem[{{O'Mara} {et~al.}(2016){O'Mara}, {Miesch}, {Featherstone}, \&
  {Augustson}}]{2016AdSpR..58.1475O}
{O'Mara}, B., {Miesch}, M.~S., {Featherstone}, N.~A., \& {Augustson}, K.~C.
  2016, Adv. Space Res., 58, 1475

\bibitem[{{Ossendrijver}(2003)}]{O03}
{Ossendrijver}, M. 2003, \aapr, 11, 287

\bibitem[{{Pencil Code Collaboration} {et~al.}(2021){Pencil Code
  Collaboration}, {Brandenburg}, {Johansen}, {Bourdin}, {Dobler}, {Lyra},
  {Rheinhardt}, {Bingert}, {Haugen}, {Mee}, {Gent}, {Babkovskaia}, {Yang},
  {Heinemann}, {Dintrans}, {Mitra}, {Candelaresi}, {Warnecke},
  {K{\"a}pyl{\"a}}, {Schreiber}, {Chatterjee}, {K{\"a}pyl{\"a}}, {Li},
  {Kr{\"u}ger}, {Aarnes}, {Sarson}, {Oishi}, {Schober}, {Plasson}, {Sandin},
  {Karchniwy}, {Rodrigues}, {Hubbard}, {Guerrero}, {Snodin}, {Losada},
  {Pekkil{\"a}}, \& {Qian}}]{2021JOSS....6.2807P}
{Pencil Code Collaboration}, {Brandenburg}, A., {Johansen}, A., {et~al.} 2021,
  The Journal of Open Source Software, 6, 2807

\bibitem[{{Proxauf}(2021)}]{2021PhDT........26P}
{Proxauf}, B. 2021, PhD thesis, Georg August University of Gottingen, Germany

\bibitem[{{Roberts}(1968)}]{1968RSPTA.263...93R}
{Roberts}, P.~H. 1968, Philosophical Transactions of the Royal Society of
  London Series A, 263, 93

\bibitem[{{Roxburgh} \& {Simmons}(1993)}]{1993A&A...277...93R}
{Roxburgh}, L.~W. \& {Simmons}, J. 1993, \aap, 277, 93

\bibitem[{{Saikia} {et~al.}(2000){Saikia}, {Singh}, {Chan}, {Roxburgh}, \&
  {Srivastava}}]{2000ApJ...529..402S}
{Saikia}, E., {Singh}, H.~P., {Chan}, K.~L., {Roxburgh}, I.~W., \&
  {Srivastava}, M.~P. 2000, \apj, 529, 402

\bibitem[{{Schrinner} {et~al.}(2012){Schrinner}, {Petitdemange}, \&
  {Dormy}}]{SPD12}
{Schrinner}, M., {Petitdemange}, L., \& {Dormy}, E. 2012, \apj, 752, 121

\bibitem[{{Schumacher} \& {Sreenivasan}(2020)}]{2020RvMP...92d1001S}
{Schumacher}, J. \& {Sreenivasan}, K.~R. 2020, Reviews of Modern Physics, 92,
  041001

\bibitem[{{Singh} {et~al.}(1995){Singh}, {Roxburgh}, \&
  {Chan}}]{1995A&A...295..703S}
{Singh}, H.~P., {Roxburgh}, I.~W., \& {Chan}, K.~L. 1995, \aap, 295, 703

\bibitem[{{Singh} {et~al.}(1998){Singh}, {Roxburgh}, \&
  {Chan}}]{1998A&A...340..178S}
{Singh}, H.~P., {Roxburgh}, I.~W., \& {Chan}, K.~L. 1998, \aap, 340, 178

\bibitem[{{Spruit}(1997)}]{Sp97}
{Spruit}, H. 1997, \memsai, 68, 397

\bibitem[{{Sreenivasan}(1984)}]{1984PhFl...27.1048S}
{Sreenivasan}, K.~R. 1984, Physics of Fluids, 27, 1048

\bibitem[{{Stein} \& {Nordlund}(1989)}]{SN89}
{Stein}, R.~F. \& {Nordlund}, A. 1989, \apjl, 342, L95

\bibitem[{{Stein} \& {Nordlund}(1998)}]{1998ApJ...499..914S}
{Stein}, R.~F. \& {Nordlund}, {\AA}. 1998, \apj, 499, 914

\bibitem[{{Stevenson}(1979)}]{1979GApFD..12..139S}
{Stevenson}, D.~J. 1979, Geophysical and Astrophysical Fluid Dynamics, 12, 139

\bibitem[{{Tremblay} {et~al.}(2015){Tremblay}, {Ludwig}, {Freytag}, {Fontaine},
  {Steffen}, \& {Brassard}}]{2015ApJ...799..142T}
{Tremblay}, P.-E., {Ludwig}, H.-G., {Freytag}, B., {et~al.} 2015, \apj, 799,
  142

\bibitem[{{Vasil} {et~al.}(2021){Vasil}, {Julien}, \&
  {Featherstone}}]{2021PNAS..11822518V}
{Vasil}, G.~M., {Julien}, K., \& {Featherstone}, N.~A. 2021, Proceedings of the
  National Academy of Science, 118, e2022518118

\bibitem[{{Vassilicos}(2015)}]{2015AnRFM..47...95V}
{Vassilicos}, J.~C. 2015, Annual Review of Fluid Mechanics, 47, 95

\bibitem[{{Vitense}(1953)}]{Vi53}
{Vitense}, E. 1953, \zap, 32, 135

\bibitem[{{Viviani} \& {K{\"a}pyl{\"a}}(2021)}]{2021A&A...645A.141V}
{Viviani}, M. \& {K{\"a}pyl{\"a}}, M.~J. 2021, \aap, 645, A141

\bibitem[{{Viviani} {et~al.}(2018){Viviani}, {Warnecke}, {K{\"a}pyl{\"a}},
  {K{\"a}pyl{\"a}}, {Olspert}, {Cole-Kodikara}, {Lehtinen}, \&
  {Brandenburg}}]{2018A&A...616A.160V}
{Viviani}, M., {Warnecke}, J., {K{\"a}pyl{\"a}}, M.~J., {et~al.} 2018, \aap,
  616, A160

\bibitem[{{Weiss} {et~al.}(2004){Weiss}, {Hillebrandt}, {Thomas}, \&
  {Ritter}}]{WHTR04}
{Weiss}, A., {Hillebrandt}, W., {Thomas}, H.-C., \& {Ritter}, H. 2004, {Cox and
  Giuli's Principles of Stellar Structure} (Cambridge, UK: Cambridge Scientific
  Publishers Ltd)

\bibitem[{{Ziegler} \& {R{\"u}diger}(2003)}]{2003A&A...401..433Z}
{Ziegler}, U. \& {R{\"u}diger}, G. 2003, \aap, 401, 433

\end{thebibliography}
